 \documentclass[twocolumn]{aastex631}

\shorttitle{Shocks in Nova Ejecta}
\shortauthors{Hachisu \& Kato}

\begin{document}


\title{A strong shock during a nova outburst: an origin of multiple
velocity systems in optical spectra and of high-energy emissions}


\author[0000-0002-0884-7404]{Izumi Hachisu}
\affil{Department of Earth Science and Astronomy, 
College of Arts and Sciences, The University of Tokyo,
3-8-1 Komaba, Meguro-ku, Tokyo 153-8902, Japan} 
\email{izumi.hachisu@outlook.jp}




\author[0000-0002-8522-8033]{Mariko Kato}
\affil{Department of Astronomy, Keio University, 
Hiyoshi, Kouhoku-ku, Yokohama 223-8521, Japan} 

%
%




\begin{abstract}
We propose a theoretical explanation of absorption/emission line systems
in classical novae based on a fully self-consistent nova explosion model.
We found that a reverse shock is formed far outside the photosphere 
($\gtrsim 10^{13}$ cm) because later-ejected mass with a faster velocity
collides with earlier-ejected matter.  Optically thick winds blow
continuously at a rate of $\sim 10^{-4} ~M_\sun$ yr$^{-1}$ near the optical
maximum but its velocity decreases toward the optical maximum and increases
afterward, so that the shock arises only after optical maximum.
The nova ejecta is divided by the shock into three parts, the outermost
expanding gas (earliest wind before maximum), shocked shell,
and inner fast wind, which respectively contribute to pre-maximum, principal,
and diffuse-enhanced absorption/emission line systems.  A large part of
nova ejecta is eventually confined to the shocked shell.  The appearance
of the principal system is consistent with the emergence of a shock.
This shock is strong enough to explain thermal hard X-ray emissions.
The shocked layer has a high temperature of
$k T_{\rm sh} \sim 1$ keV 
$\times ((v_{\rm wind} - v_{\rm shock})/{\rm 1000 ~km~s}^{-1})^2
={\rm 1 ~keV}\times ((v_{\rm d} - v_{\rm p})/{\rm 1000 ~km~s}^{-1})^2$,
where $v_{\rm d} - v_{\rm p}$ is the velocity difference
between the diffuse-enhanced ($v_{\rm d}$)
and principal ($v_{\rm p}$) systems.
We compare a 1.3 $M_\sun$ white dwarf model
with the observational properties of the GeV gamma-ray detected
classical nova V5856 Sgr (ASASSN-16ma) and discuss what kind of novae
can produce GeV gamma-ray emissions.
\end{abstract}




\section{Introduction}
\label{introduction}

One of the long-standing problems in novae is 
the origin of different velocity systems of absorption/emission
lines in nova spectra. 
\citet{mcl42} firstly classified nova spectra into three systems:  
(i) the pre-maximum, (ii) principal, and (iii) diffuse-enhanced 
absorption/emission line systems.
The pre-maximum system is observed only in the pre-maximum phase 
and up to a few to several days after the maximum.
The principal and diffuse-enhanced systems show different velocities,   
both increase with time \citep[see Tables 3 and 5 of ][]{mcl43}.
The principal system becomes dominant after the optical maximum. 
These multiple velocity systems indicate 
multiple mass ejections with different velocities. 
However, no theoretical work has explained
such multiple mass ejections within the framework of 
one-dimensional (1D) approximation.  

In the last decades, hard X-ray and GeV gamma-ray emissions have 
been observed in classical novae. 
Hard X-rays were detected in a middle phase of a nova outburst, i.e.,
from a few $t_3$ times to several $t_3$ times after the optical maximum
\citep[e.g.,][]{llo92ob, bal98ko, muk01i}, where $t_3$ time is 
the 3 mag decay time from the optical peak. 
GeV gamma-ray emissions were observed in an early phase of a nova, 
from the post-maximum phase, and continues a few tens of days 
\citep[e.g.,][]{abd10, ack14, li17mc, gor21ap}.

These high-energy (hard X-rays and GeV $\gamma$-rays)
emissions are considered to originate from strong shocks
between shells ejected with different velocities
\citep{cho14ly, met15fv, mar18dj}.
If the inner shell (later ejected) has a larger velocity than that of
the outer shell (earlier ejected), the inner one can catch up with the 
outer one and forms a shock wave \citep[e.g.,][]{muk19s, ayd20ci, ayd20sc}. 
The assumption of multiple shell ejections is the key of this idea. 

In this way, multiple shell ejections were suggested
from both optical and high-energy emissions from novae.
There is, however, no unified theory that naturally explains all
these different wavelength observations based on nova explosion models
\citep[see][for a recent review]{cho21ms}.

From the theoretical point of view, many numerical calculations 
have been presented from the early thermonuclear runaway to the  
extended phase of nova outbursts \citep[e.g.,][]{
pri92k, pri95k, epels07, sta09ih, den13hb, chen19, kat22sha, kat22shb}. 
These works clarified that mass ejection is continuous,
and no multiple distinct mass ejections occur. 
Also no shock arises at the thermonuclear runaway 
(see section \ref{section_shock} for more detail).

Thus, we have to bridge the gap between these theoretical understandings
of continuous mass loss and distinct multiple mass ejections.  
The aim of this paper is to present a natural explanation 
for a shock formation during nova outburst and multiple velocities of
ejecta based on our theoretical nova evolution models \citep{kat22sha}. 



This paper is organized as follows. 
First we summarize the current common understanding 
of a shock in nova theory (section \ref{section_shock}). 
Then, we introduce our nova model in
section \ref{section_nova_model}, and  
clarify that a shock is formed far outside the photosphere
in section \ref{internal_shock_formation}. 
The origin of absorption/emission line systems is presented
and discussed in section \ref{three_velocity_systems}. 
High-energy (hard X-rays and GeV $\gamma$-rays) emissions from a shock
are separately discussed in section \ref{section_emission}. 
A discussion follows in sections 
\ref{super_eddington}-\ref{calm_violent_ejection}.
Especially,
we compare the present results with the observational properties of 
the classical nova V5856 Sgr (ASASSN-16ma)
in section \ref{comparison_asassn16ma_v5856sgr}. 
The conclusion follows in section \ref{conclusions}.

\section{Shock formation outside the photosphere}
\label{section_model}

\subsection{Where is a Shock Arising?}
\label{section_shock}

Nova outbursts have been studied since the 1970s by various authors.  
Among those people, \citet{sta72tsk} calculated a strong shock 
associated with thermonuclear runaway. 
\citet{pri86} wrote that a nova outburst experiences 
massive shock ejection. 
\citet{pri92k} later wrote ``we find continuous 
mass loss instead of shock ejection. These differences may 
be the result of our more accurate treatment of mass loss.''
Since then, many numerical calculations have been presented
\citep[e.g.,][]{pri95k, epels07, sta09ih, den13hb, chen19}, 
but no one detected a shock wave. 

\citet{kat22sha} explained why a shock wave does not arise
around the nuclear burning region. 
The timescale  of nuclear energy generation ($\sim 100$ s) 
is larger (longer) than the hydrodynamic timescale ($\sim 0.4$ s). 
A huge amount of nuclear energy is generated, which is once absorbed by 
the layer around nuclear burning region and then will be released
in a later phase with a long timescale. 
During the nova outburst, the white dwarf (WD) envelope quickly expands 
to a giant size, but no shock arises below the photosphere
\citep[e.g.,][]{den13hb, kat22sha, kat22shb}. 
Thus, the statement ``no shock arises during/after 
thermonuclear runaway'' is the current understanding in nova theory.


\begin{figure}
\gridline{\fig{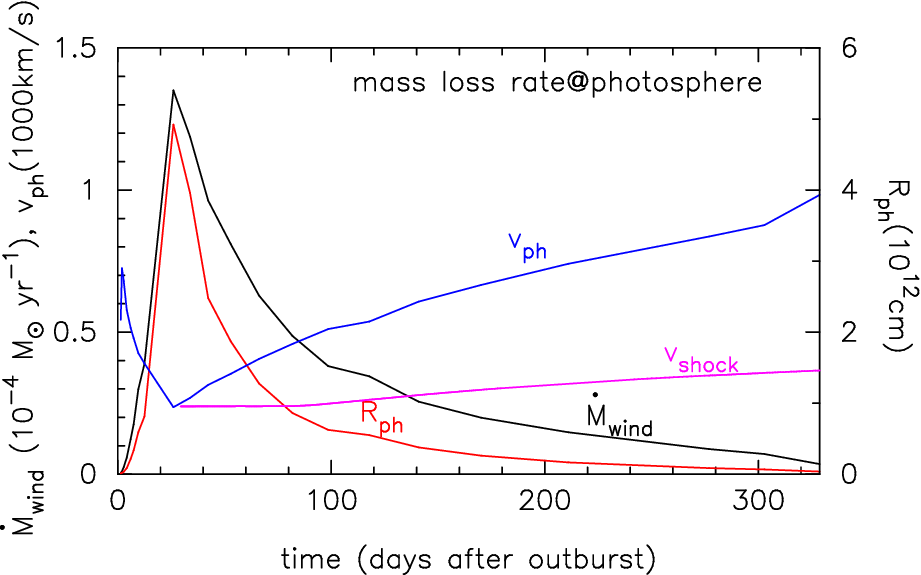}{0.47\textwidth}{(a)}
          }
\gridline{
          \fig{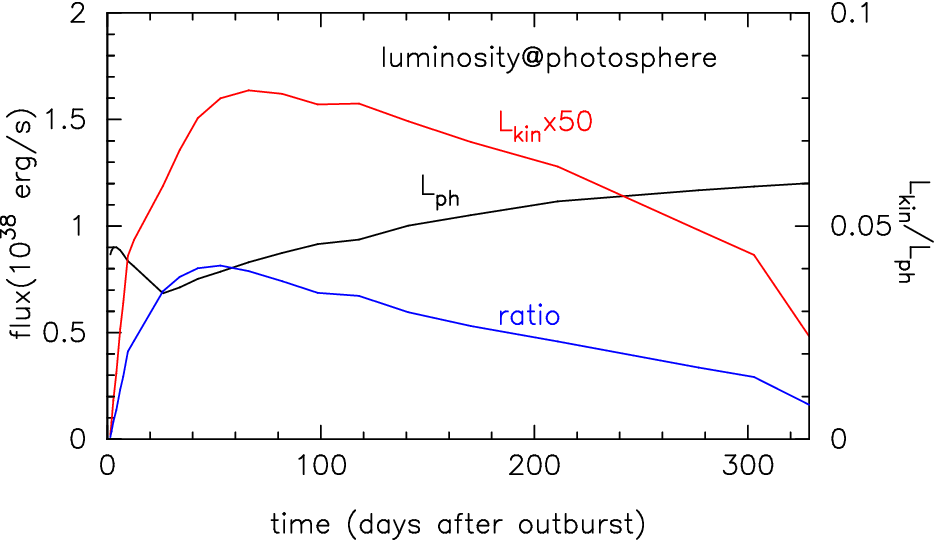}{0.47\textwidth}{(b)}
          }
\caption{
Various photospheric properties of a 1.0 $M_\sun$ WD model \citep{kat22sha}.  
{\bf (a)} The wind mass-loss rate (solid black line, 
labeled ${\dot M}_{\rm wind}$), wind velocity (blue, $v_{\rm ph}$), 
and photospheric radius (red, $R_{\rm ph}$) are plotted
against time (days after outburst).
  We also add the shock velocity
(magenta, $v_{\rm shock}$).  The shock is located far outside
the photosphere (see Figure \ref{shock_density_pp_interaction}a
for the position of the shock).
{\bf (b)} The photospheric luminosity, $L_{\rm ph}$, is plotted against time
(days after outburst).  We also add the kinetic energy of the wind
at the photosphere, $L_{\rm kin}$, which is multiplied by 50.
The ratio of $L_{\rm kin}$ to $L_{\rm ph}$ is also plotted.
The kinetic energy does not exceed 5\% of the photospheric luminosity.
\label{mdot_radius_velocity}}
\end{figure}


\begin{figure}
\epsscale{1.15}
\plotone{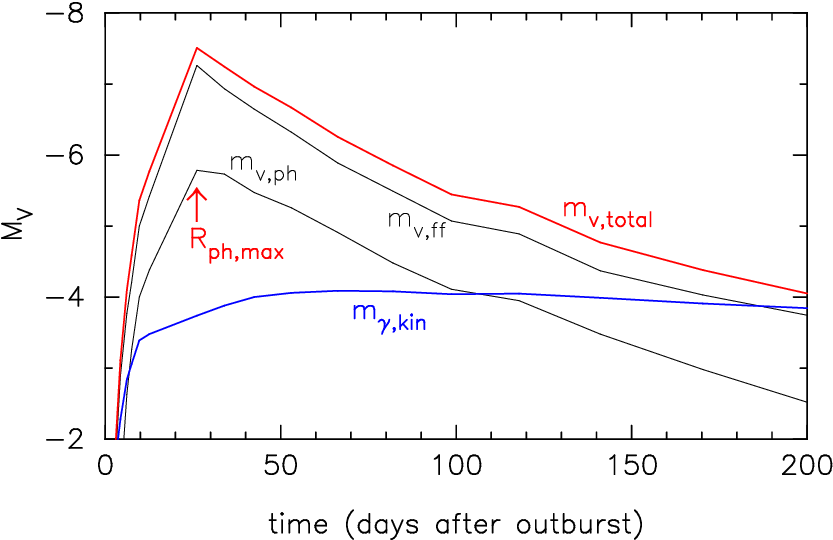}
\caption{
Three absolute $V$ magnitudes,
$m_{V, \rm total}$, $m_{V, \rm ff}$,
and $m_{V, \rm ph}$  are plotted against time (days after outburst).
The total $V$ magnitude is the sum of the free-free emission component
($m_{V, \rm ff}$) and the photospheric blackbody component
($m_{V, \rm ph}$).  The optical maximum in the $m_{V, \rm total}$ magnitude 
corresponds to the maximum expansion of the photosphere
(labeled $R_{\rm ph, max}$).  No circumstellar or interstellar extinctions
are included. We also add the kinematic component (kinetic energy
of wind) of $m_{\gamma, \rm kin}$.  The three $V$ magnitudes of
$m_{V, \rm total}$, $m_{V, \rm ff}$, and $m_{V, \rm ph}$ are calibrated,
where the distance $d=10$ pc is assumed,
but only the $m_{\gamma, \rm kin}$ magnitude is scale-free, not calibrated.
\label{opt_gamma_ray_abs_magnitudes}}
\end{figure}

\subsection{A Self-consistent Nova Model}
\label{section_nova_model}

Although no shock arises inside the photosphere,
a shock could be generated {\it outside} the photosphere. 
Matter is ejected through the photosphere.
Such matter has usually been removed from the code,
mainly because of difficulty of calculation 
in Henyey-type codes \citep[see, e.g.,][for details]{kat17palermo}.

In the present paper, we use the nova model 
calculated by \citet{kat22sha}. 
They obtained internal structure from the WD center to the photosphere, 
consistently including wind acceleration of 
radiation-pressure gradient (so-called optically thick wind).   
This model is for a $1.0~M_\sun$ WD with
the mass accretion rate to the WD before the outburst of
${\dot M}_{\rm acc}= 5\times 10^{-9}$ $M_\sun$ yr$^{-1}$, 
a typical parameter for classical novae. This is only the  
fully self-consistent model of classical nova calculation
\citep[see also][for another self-consistent model, but for recurrent novae
of 1.2 $M_\sun$ and 1.38 $M_\sun$ WDs]{kat17sh}. 

Figure \ref{mdot_radius_velocity}a shows the photospheric 
radius and wind mass-loss rate of the nova outburst model mentioned above.
The thermonuclear runaway starts at $t=0$, and then
the hydrogen-rich envelope begins to expand. 
The wind (optically thick wind) is accelerated at $t=1.05$ day. 
The photospheric radius reaches the maximum value of 
$R_{\rm ph}\sim 5\times 10^{12}$ cm ($\approx 71$ $R_\sun$) 
at $t=26$ day, where the 
photospheric temperature is $T_{\rm ph}\sim 8000$ K 
(not shown in Figure \ref{mdot_radius_velocity}). 
The wind mass loss rate also reaches the maximum,  
${\dot M}_{\rm wind}\sim 1.4\times 10^{-4}$ $M_\sun$ yr$^{-1}$. 
After that,
the photospheric radius decreases and the temperature increases with 
time because the envelope loses its mass owing to optically thick winds. 

This figure also shows the velocity of outgoing matter at the photosphere.  
The velocity is smaller than a typical absorption line
velocity of fast novae ($1000$--$2000$ km s$^{-1}$) 
but consistent with that of slow and very slow novae 
($200$--$800$ km s$^{-1}$)\footnote{The nova speed class is defined
by $t_3$ or $t_2$ (days of 3 or 2 mag decay from the optical maximum).
For example, 
very fast novae ($t_2 \le 10$ day), 
fast novae ($11 \le t_2 \le 25$ day), 
moderately fast nova ($26 \le t_2 \le 80$ day), 
slow novae ($81 \le t_2 \le 150$ day), 
and very slow novae ($151 \le t_2 \le 250$ day), 
which are defined by \citet{pay57}.}. 
The velocity of wind $v_{\rm ph}$ decreases toward the optical maximum
and then increases afterward (blue line in Figure 
\ref{mdot_radius_velocity}a).  This trend in the photospheric velocity
$v_{\rm ph}$ is commonly observed in many novae.
For example, \citet{ayd20ci} showed temporal variations of
the velocities in the H$\alpha$ or H$\beta$ P Cygni profiles of four novae, 
V906 Car, V435 CMa, V549 Vel, and V5855 Sgr in their Figure 11. 
Among the four novae, three novae, V906 Car, V435 CMa, and V5855 Sgr,
show slow velocities of $\sim 200-250$ km s$^{-1}$ near the maximum,
which are consistent with our model of 
$v_{\rm ph}\sim 235$ km s$^{-1}$ at maximum expansion of
the photosphere (see Figure \ref{mdot_radius_velocity}a).

Figure \ref{mdot_radius_velocity}b shows the temporal variations of
the photospheric luminosity $L_{\rm ph}$, 
kinetic energy flux of wind
$L_{\rm kin}\equiv {\dot M}_{\rm wind}v_{\rm ph}^2/2$,
and the ratio of $L_{\rm kin}/L_{\rm ph}$.
The photospheric luminosity decreases toward the optical 
maximum and increases after that.
The optical ($V$ band) maximum of this model corresponds to
the maximum expansion of the photosphere and, at the same time,
the maximum wind mass-loss rate (see Figure 
\ref{opt_gamma_ray_abs_magnitudes} for the $V$ magnitude of this model,
the details of which are presented in section \ref{super_eddington}).
This is due to the energy balance in the envelope 
in which a larger part of the diffusive luminosity 
is consumed to lift up the wind matter, so 
the photospheric luminosity is smaller when 
the mass-loss rate is larger 
\citep[see Figure 13 of][for the energy balance
in the nova envelope]{kat22sha}.

The kinetic energy flux of the wind reaches 
$L_{\rm kin}\approx 3.2\times 10^{36}$ erg s$^{-1}$ 
just in the post-maximum phase, about 5\% of 
the photospheric luminosity at the optical maximum.
This small kinetic energy comes from energy balance in the envelope,
in which the wind is accelerated owing to
the radiation-pressure gradient deep inside the photosphere. 
It should be noted that the radiative energy is consumed to lift up
the massive wind, and the residual is not enough to accelerate the wind
to much higher velocities \citep[see][]{kat22sha}.  

Even if a strong shock is formed above (outside) the photosphere 
and all the kinetic energy flux is converted to the optical 
luminosity at the shock, the luminosity of shock
may not be larger than a few percent of the photospheric luminosity. 
Therefore, it is not enough to explain the super-Eddington luminosity
as long as \citet{kat22sha}'s model is concerned.

\subsection{Formation of a Strong Shock Outside the Photosphere}
\label{internal_shock_formation}

\begin{figure}
\gridline{\fig{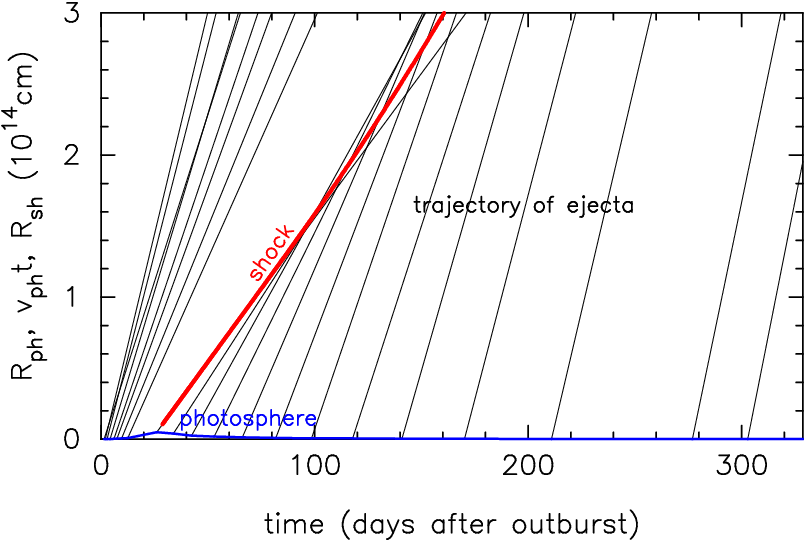}{0.47\textwidth}{(a)}
          }
\gridline{
          \fig{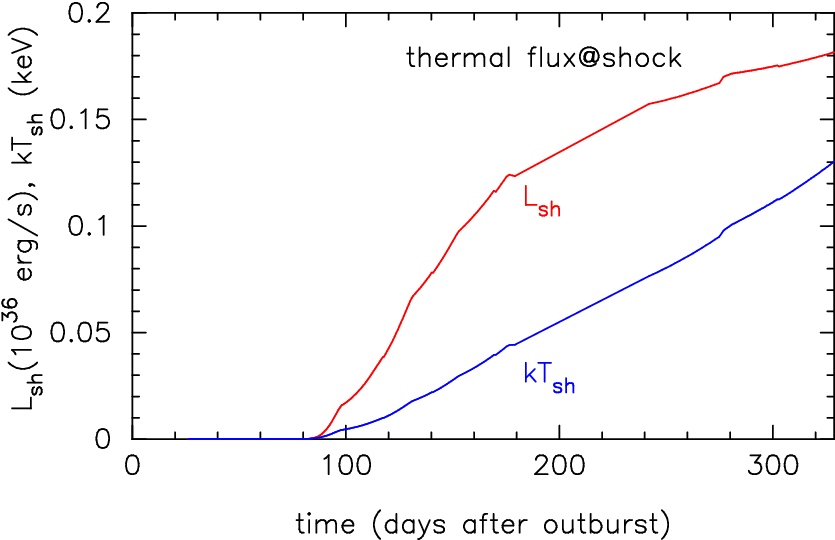}{0.47\textwidth}{(b)}
          }
\gridline{
          \fig{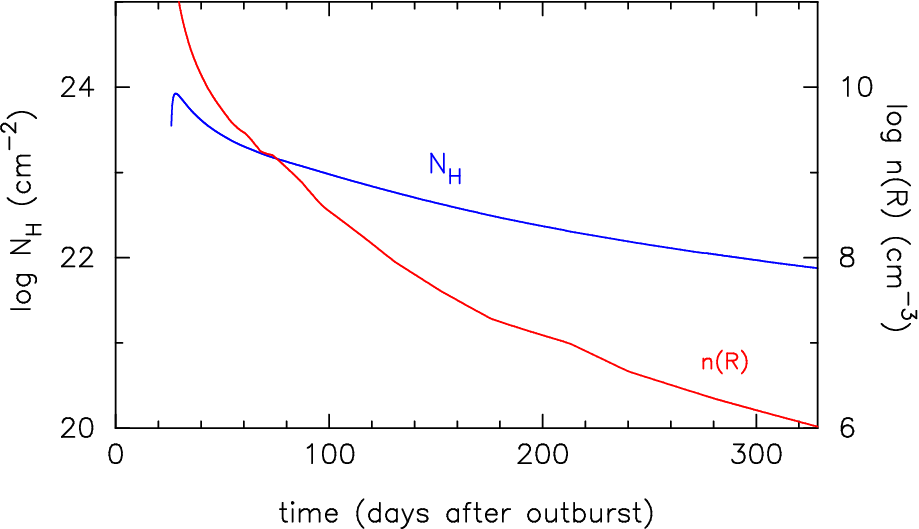}{0.5\textwidth}{(c)}
          }
\caption{
{\bf (a)} Trajectories (thin black line) of the winds ejected
from the model in Figure \ref{mdot_radius_velocity}.
We also add the position of the strong shock (thick red line).
The thick blue line shows the position of the photosphere. 
{\bf (b)} Temporal variation in the luminosity $L_{\rm sh}$ 
(solid red line) and temperature k$T_{\rm sh}$ (solid blue line)
of shocked matter.
{\bf (c)} Temporal variation in the hydrogen column density $N_{\rm H}$
(blue line) behind the shock and the number density $n(R)$
(red line) just in front of the shock ($R=R_{\rm sh}$).
\label{shock_density_pp_interaction}}
\end{figure}

Figure \ref{mdot_radius_velocity}a shows the velocity of the wind at the 
photosphere, which decreases before the maximum expansion of
the photosphere (optical maximum) and increases after that. 
Assuming a ballistic motion of a fluid particle, 
i.e., the velocity of gas is constant outside the photosphere, 
we plot the trajectory of wind in Figure 
\ref{shock_density_pp_interaction}a. 
Before the optical maximum, the photospheric velocity decreases 
with time, so each locus departs from the others. 
After the optical maximum, on the other hand, 
the trajectory of the wind converges, i.e., the wind ejected later
catches up with the previously ejected matter. 
Thus, matter will be compressed, which causes a strong shock wave. 
This occurs after the optical maximum. 
The temporal variation in the shock velocity 
is shown by the solid magenta line
in Figure \ref{mdot_radius_velocity}a.

Theoretically the shock wave arises outside the photosphere.   
However, observationally it is called the ``internal shock,'' 
which simply means that there is a shock in the ejecta.  

We calculated the location of the shock, which is shown by
the thick red line in Figure \ref{shock_density_pp_interaction}a. 
The location of the shock moves outward and 
the separation from the photosphere increases with time. 
Here, we assume that the shocked region cools down in a shorter timescale
than a dynamical timescale \citep[isothermal shock; see, e.g.,][]{met14hv,
met15fv, mar18dj}, and the shocked layer is geometrically thin.
We can derive the velocity of the shock ($v_{\rm shock}$) 
from the mass ($M_{\rm shell}$) and momentum 
($M_{\rm shell} v_{\rm shock}$) of this thin shell \citep{cho21ms}.
The mass of the shocked shell ($M_{\rm shell}$) increases with time and 
reached about 90\% of the total ejecta mass, i.e., $M_{\rm shell}\sim
2.7 \times 10^{-5} ~M_\sun$.  In other words,
a large part of nova ejecta is eventually confined to the shocked shell.


\subsection{Short Summary}
\label{shortl_summary_shock_formation}

To summarize, we first show a shock generation far outside
the nova photosphere, without any arbitrary assumptions on multiple
mass ejections.  
\citet{kat22sha}'s calculation fully includes thermonuclear runaway, 
envelope expansion, and wind acceleration in a self-consistent manner.
As a result, they obtained temporal change in the wind velocity at 
the photosphere.  The velocity of the wind
decreases toward the maximum expansion of the photosphere
and turn to increase after that.
The wind acceleration is closely related with both the envelope structure 
and radiative transfer deep inside the photosphere.
So, this trend in the temporal change in the wind velocity is common
among other nova outbursts even if they have different parameters.
Thus, we may conclude that a shock wave generation far outside 
the photosphere is common among nova outbursts, and it occurs
after the maximum expansion of the photosphere.
The epoch at the maximum expansion of the photosphere coincides with that 
of the optical maximum (see Figure \ref{opt_gamma_ray_abs_magnitudes}).


\begin{figure*}
\epsscale{0.85}
\plotone{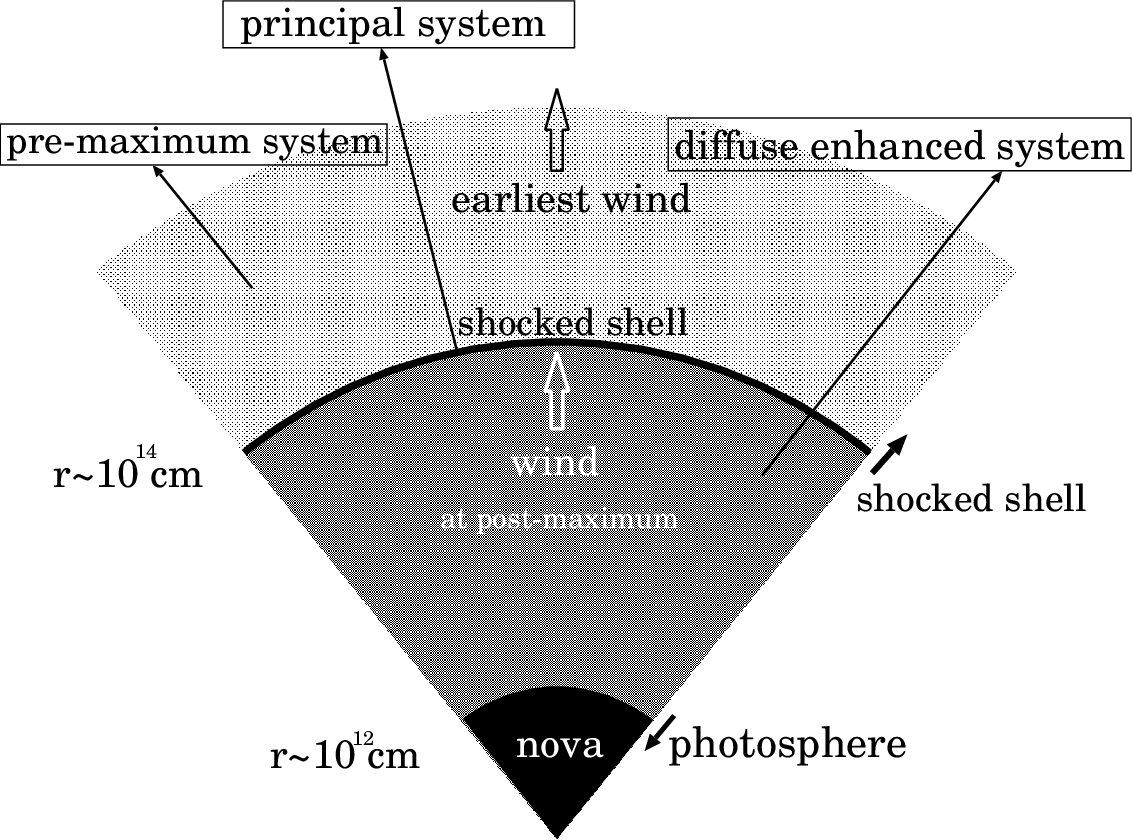}
\caption{
Cartoon for our shocked ejecta model.  The shock is formed 
(at $R_{\rm sh}\sim 10^{13}-10^{14}$ cm) far outside
the photosphere ($R_{\rm ph} \sim 10^{12}$ cm).
The shocked layer is geometrically thin, and the whole ejecta is
divided into three parts, outermost expanding gas (earliest wind),
shocked thin shell, and inner wind.
These three parts contribute to pre-maximum, principal, and
diffuse-enhanced absorption/emission line systems
\citep{mcl42, mcl60}, respectively.
The shocked shell emits thermal hard X-rays.
\label{wind_shock_config}}
\end{figure*}

\section{Three Velocity Systems}
\label{three_velocity_systems}

Figure \ref{wind_shock_config} shows a schematic illustration
of an expanding nova shell.  The shock forms a thin shell,
which divides the ejecta into three parts.
Correspondingly, we have three velocity components:
(1) outermost expanding gas (earliest wind),
(2) shocked shell, and
(3) wind in the post-maximum phase, i.e., after the maximum expansion
of the photosphere (post-maximum wind),
as illustrated in Figure \ref{wind_shock_config}.
These three have velocities of $v_{\rm early}\sim 725$--$235$
km s$^{-1}$, $v_{\rm shock}\sim 240$ km s$^{-1}$, 
and $v_{\rm wind}\sim 245$--$450$ km s$^{-1}$, respectively,
at $t\sim 80$ day in Figures \ref{mdot_radius_velocity}a
and \ref{shock_density_pp_interaction}a.

The total mass of the shocked shell increases with time,
about $M_{\rm shell}\sim 1.2\times 10^{-5} ~M_\sun$ at $t=80$ day,
and finally reaches
$M_{\rm shell}\sim 2.7\times 10^{-5} ~M_\sun$ at $t=350$ day.
The total ejected mass is about $M_{\rm ej}\sim 3.0\times 10^{-5} ~M_\sun$
in \citet{kat22sha}'s model. 
The other mass of $\Delta M= M_{\rm ej}- M_{\rm shell} \sim 
0.3\times 10^{-5} ~M_\sun$ forms early expanding matter outside
the shocked shell, as shown in Figures \ref{shock_density_pp_interaction}a
and \ref{wind_shock_config}. 

We propose that these three parts of ejecta
correspond, respectively, to those that make pre-maximum, principal, and
diffuse-enhanced absorption/emission line systems.  
In the followings we examine these correspondences in more detail. 

\subsection{Various Features of Principal and Diffuse-enhanced Systems}
\label{various_feature_line_systems}

\citet{mcl42} pointed out various features of
the principal absorption/emission line system in comparison with other 
velocity systems. 

\subsubsection{Emergence of Principal System}
\label{emergenced_principal_system}

The principal system becomes dominant a few days after the optical maximum. 
This property is consistent with our model that a shock emerges after the 
optical maximum.  \citet{mcl42} also wrote ``its displacement is greater
than that of the pre-maximum spectrum.'' The pre-maximum winds have 
a wide range of velocities, the lowest tail of which is the lowest wind
velocity at the optical maximum.  So, this can be explained if the velocity
of the shocked layer could quickly increase higher than that of
the slowest pre-maximum (earlier) winds as can be imagined from Figure
\ref{mdot_radius_velocity}a and Figure \ref{shock_density_pp_interaction}a.

\subsubsection{Velocities in Principal and Diffuse Enhanced Systems}
\label{velocities_principal_enhanced_systems}

\citet{mcl43} listed velocities of pre-maximum, principal,
and diffuse-enhanced systems in several novae 
(in their Tables 3 and 5). These velocities are different from
nova to nova, but trends of our three velocity components are
qualitatively consistent with the absorption/emission line systems
\citep{mcl42}. 
For example, the velocities of diffuse-enhanced systems are
about double those of principal systems \citep{mcl42}.
In our model, the velocity of post-maximum wind
($v_{\rm ph}\sim 500$ km s$^{-1}$) is about double the
velocity ($v_{\rm shock}\sim 250$ km s$^{-1}$) of the shock at $t=80$ day. 
This quantitative agreement of the ratio (about double) seems to support
our interpretation on the assignment of three parts of ejecta to three 
absorption/emission line systems.

\subsubsection{Dependence on the Nova Speed Class}
\label{dependence_on_nova_speed_class}

The velocities of principal absorption systems show a large variation
between $-300$ km s$^{-1}$ and $-1500$ km s$^{-1}$
\citep[e.g., Table 3 of ][]{mcl43}.  \citet{mcl60} found an
empirical relation between the velocity of principal absorption
system $v_{\rm p}$ (km s$^{-1}$) and the nova speed class $t_3$ or $t_2$,
i.e., 
\begin{equation}
\log v_{\rm p} = 3.7 - 0.5 \log t_3 = 3.57 - 0.5 \log t_2. 
\label{vp_vs_t3}
\end{equation}

Our theoretical nova model produces an optical $V$ light curve 
as shown in Figure \ref{opt_gamma_ray_abs_magnitudes}, in 
which we read $t_3\approx 145$ day (or $t_2\approx 73$ day).
This corresponds to the moderately fast nova class \citep{pay57}, but
close to the slow nova class. 
Substituting this value into equation (\ref{vp_vs_t3}),
we obtain $v_{\rm p}\approx 400$ km s$^{-1}$
for the principal system of our model. 
The $v_{\rm shock}$ of our model increases from 240 km s$^{-1}$
at $t= 40$ day to 360 km s$^{-1}$ at $t= 300$ day
as seen in Figure \ref{mdot_radius_velocity}a,
being broadly consistent with $v_{\rm p}\sim 400$ km s$^{-1}$.

For the velocity of diffuse-enhanced absorption system
$v_{\rm d}$ (km s$^{-1}$), 
\citet{mcl60} also obtained an empirical relation of 
\begin{equation}
\log v_{\rm d} = 3.81 - 0.4 \log t_3 = 3.71 - 0.4 \log t_2.
\label{vd_vs_t3}
\end{equation}
We have $v_{\rm d}\approx 880$ km s$^{-1}$ for $t_3\approx 145$ day.
The $v_{\rm ph}$ of our model increases
from 240 km s$^{-1}$ at $t= 40$ day to 700 km s$^{-1}$ at $t= 300$ day
as seen in Figure \ref{mdot_radius_velocity}a. 
This is a bit small but broadly consistent with 
$v_{\rm d}\approx 880$ km s$^{-1}$.

In this way the empirical relations of $v_{\rm p}$-$t_3$ 
and $v_{\rm d}$-$t_3$ are broadly consistent with our model,
which suggests our interpretations being reasonable.  
In what follows, we regard that $v_{\rm p}= v_{\rm shock}$ and
$v_{\rm d}= v_{\rm ph}$.

\subsection{Origin of Hybrid He/N--\ion{Fe}{2}--He/N Novae}
\label{hybrid_he/n_fe2_novae}

Novae are divided into two classes, He/N or \ion{Fe}{2}, from 
their spectral features near the optical maximum \citep{wil92}.
He/N novae have larger expansion velocities and a higher level of
ionization.  \ion{Fe}{2} type novae evolve more
slowly, and have a lower level ionization, and show P Cygni components.
\citet{wil92} regarded that the He/N novae are different from 
the \ion{Fe}{2} novae in the aspect of mass ejection.
He wrote ``the spectra can be interpreted in terms of a two-component
gas consisting of a discrete shell and a continuous wind,
in which the narrower \ion{Fe}{2} spectrum is formed in the wind,
while the broader He/N spectrum is formed in the shell
ejected at maximum light.''  His outer shell/inner wind structure
is similar to our shock model in Figure \ref{wind_shock_config},
but the velocities of shell and wind are reversed.
In our model, the shocked shell has a slower velocity, while the inner
winds have much faster velocities.

\citet{tan11nf} found that the spectral type of V5558 Sgr evolved from
the He/N type toward the \ion{Fe}{2} type during a pre-maximum halt
and then toward the He/N type again.
The recurrent nova T Pyx also shows a similar transition
to V5558 Sgr, that is,
from He/N to \ion{Fe}{2}, and then to He/N again \citep{izz12ed, ede14}.

In \citet{kat22sha}'s model, the nova has a high photospheric
temperature and tenuous wind (low $\dot{M}_{\rm wind}$) having
higher velocities in an early phase as shown in Figure
\ref{mdot_radius_velocity}a ($T_{\rm ph}$ is not shown).
The photospheric temperature decreases from 100,000 K to 10,000 K
in this early phase (see Figure 2 of \citet{kat22sha}). 
These aspects are suitable for a formation of He/N nova.

Then the temperature decreases to $T_{\rm ph}\sim 8000$ K
at the optical maximum,
which is as low as typical nova temperatures at optical maximum.
The wind-mass loss rate becomes $\dot{M}_{\rm wind}\sim
(1-2)\times 10^{-4} ~M_\sun$ yr$^{-1}$ near the maximum,
as massive as the mass-loss rate for typical novae. 
The mass outflow velocity is also as low as 
$v_{\rm ph}\sim (200$--$300)$ km s$^{-1}$ near the maximum.
These properties are consistent with a formation site of 
narrow line \ion{Fe}{2} features \citep{wil12}.

In a later phase (post-maximum), the photosphere shrinks and
the photospheric temperature increases.
The wind mass-loss rate decreases, and its velocity increases.
These features are again favorable for a formation of He/N nova. 
Thus, the evolution of He/N -- \ion{Fe}{2} -- He/N is naturally explained.

\subsection{Transient Heavy-element Absorption Features in Novae}
\label{thea_features_novae}

Relatively narrow velocity dispersion ($\sim 30$
-- $300$ km s$^{-1}$) absorption components were observed in novae
near the optical maximum, both before and after the maximum \citep{wil08md}.
\citet{wil08md} dubbed these narrow absorption line features the
Transient Heavy-Element Absorption (THEA) system.
They concluded that the THEA gas originates before the outburst and 
most likely comes from the secondary star.
As already mentioned in section \ref{section_nova_model}, our model 
as well as the three novae, V906 Car, V435 CMa, and V5855 Sgr
\citep[see Figure 11 of ][]{ayd20ci}, show slow absorption radial
velocities of $\sim 200$--$250$ km s$^{-1}$ near the maximum.
These low-velocity components could form THEA features near the maximum,
that is, both in pre-maximum and post-maximum phases
because dense winds, $\rho = \dot{M}_{\rm wind}/4 \pi r^2 v_{\rm ph}$,
as well as the relatively low photospheric temperature of 
$T_{\rm ph}\sim 8000$ K near optical maximum \citep{kat22sha}
are favorable to the formation of the THEA systems.

\citet{ayd20ci} found that the THEA lines of V906 Car exhibit a slow
(pre-maximum) component and an intermediate component, which are 
essentially identical to the velocities and evolution of
the \ion{H}{1}, \ion{Fe}{2}, \ion{O}{1}, and \ion{Na}{1} lines,
and concluded that the THEA absorptions originate from the same body
of gas responsible for the P Cygni profiles in prominent lines.
Therefore, we regard that the THEA lines are essentially the same as
those of pre-maximum or principal system proposed by \citet{mcl42}.  
We further argued that the THEA systems are created by the early/earliest
wind (pre-maximum system) or shocked shell (principal system)
in Figure \ref{wind_shock_config}.

\citet{wil08md} obtained the excitation temperatures,
 $T_{\rm exc} \sim$ 8573 K (\ion{Sc}{2}), 9766 K (\ion{Ti}{2}),
and 11822 K (\ion{Fe}{2}), for the primary THEA system of Nova LMC 2005.
They also estimated the mass of THEA system to be as much as
$\sim 10^{-5}~M_\sun$.  The excitation temperature is consistent with
the photospheric temperature of $T_{\rm ph}\sim$ 8000--10000 K near the 
optical maximum in our model \citep{kat22sha}.  The ejecta mass reaches 
$M_{\rm ej}\sim 0.3 \times 10^{-5} ~M_\sun$ at the optical maximum
or increases to $M_{\rm ej}\sim 1.5 \times 10^{-5} ~M_\sun$ at $+54$ days
after the optical maximum in our shock model as already obtained in
section \ref{three_velocity_systems}.  Here, the ejecta mass
is calculated to be $M_{\rm ej} = M_{\rm early} + M_{\rm shell}
= 0.3 \times 10^{-5} ~M_\sun + 1.2 \times 10^{-5} ~M_\sun =
1.5 \times 10^{-5} ~M_\sun$ at $t=80$ day ($+54$
days after the optical maximum), where $M_{\rm early}$ is the total 
wind mass emitted before optical maximum.  Thus, the mass of the 
THEA system is also consistent with \citet{wil08md}'s estimate.

\citet{wil10m} wrote ``the column density of the absorbing systems,
assuming roughly solar abundances, is of the order of $\sim 10^{23}$
cm$^{-2}$ for the THEA systems.''
Our shocked shell model in Figure \ref{shock_density_pp_interaction}c
shows $N_{\rm H}\gtrsim 10^{23}$ cm$^{-2}$ for about 50 days after the 
optical maximum, which is consistent with \citet{wil10m}'s estimate
for the THEA systems.

\citet{wil10m} also added ``since THEA absorption is observed in 80\%
of novae the gas must exist along most lines of sight, i.e., have a
roughly spherical distribution'', which is also consistent with our
spherically symmetric shocked shell (or pre-maximum wind) model.

\citet{ara16ks} reported the high-dispersion spectroscopic observation 
(with resolving power of $R \sim 72,000$) of
the classical nova V2659 Cyg (Nova Cyg 2014) at 33.05 days after the  
optical $V$ maximum.  Their spectra showed two distinct blueshifted
absorption systems originating from \ion{H}{1}, \ion{Fe}{2}, \ion{Ca}{2},
etc. The velocities of the absorption systems are $-620$ km s$^{-1}$,
and from $-1100$ to $-1500$ km s$^{-1}$, which are dubbed the low-velocity
components (LVCs) and high-velocity components (HVCs), respectively.
They wrote ``the slow radial velocities and the sharp absorption line
profiles of the LVCs are consistent with those
of THEA reported by \citet{wil08md} and \citet{wil10m}.''
Thus, we may conclude that the LVCs are essentially the same as 
those of pre-maximum or principal systems defined by \citet{mcl42}.

\citet{ara16ks} also found that the absorption line profiles of the LVCs
show strong asymmetries.  Their Figure 5 displays
enlarged views of the LVCs of \ion{Fe}{2} (42) and H$\delta$.
For all line profiles, the red-side wings of the LVCs
are narrower (HWHM  $30-35$ km s$^{-1}$) than those of the blue-side
wings (HWHM  $50-70$ km s$^{-1}$).  Such a line profile (blueward extension)
can be explained in our shock model (see Figures \ref{mdot_radius_velocity}a,
\ref{shock_density_pp_interaction}a, and \ref{wind_shock_config}),
because the earliest/earlier winds (before the optical maximum) have 
larger velocities than that of the shocked shell, 
i.e., $v_{\rm early} > v_{\rm shock}$.
The main deep absorption is formed by the shocked shell and 
the blueward extensions come from the absorption of wind 
outside the shocked shell (earliest wind in Figure 
\ref{wind_shock_config}).  


\citet{taj16sn} reported similar P Cygni absorption profiles of
HVCs and LVCs of $^7$\ion{Be}{2}, \ion{Fe}{2} (42), and H$\gamma$
for V5668 Sgr (at $+$69 days after the optical maximum) and V2944 Oph
($+$80 days).  The velocities of the absorption systems are 
$-786$ km s$^{-1}$ (LVC), and from $-1350$ to $-2200$ km s$^{-1}$ (HVCs)
for V5668 Sgr (with the resolution of $R\sim 60,000$),  
$-878$ km s$^{-1}$ (LVC) and from $-1300$ to $-2000$ km s$^{-1}$ (HVCs)
for V2944 Oph ($R\sim 45,000$),
although the $^7$\ion{Be}{2} $\lambda\lambda$3130.583 and 
3131.228 doublet lines coalesce.    
The LVCs are essentially the same as the THEA systems, which correspond
to the principal systems while the HVCs correspond to the diffuse-enhanced
systems as mentioned above.  
$^7$Be ions have been created during the thermonuclear runaway (TNR) of
these novae and decayed to form $^7$Li within a short period (with a
half-life time of 53.22 days).  This strongly suggests that the absorbing
materials in both HVCs and LVCs consist of products of TNR;
that is, the THEA systems are not originated from circumstellar matter
ejected before outburst from the companion star
as \citet{wil08md} proposed.

To summarize, the THEA systems are essentially the same as those of
LVCs or intermediate-velocity components of \citet{ayd20ci}
and LVCs of \citet{ara16ks} and \citet{taj16sn}.  These velocity systems
are created by the early/earliest wind (pre-maximum systems)
or shocked shell (principal systems) near the optical maximum.
In our shock model, the asymmetry of LVCs (same as THEA)
are naturally explained by absorption of the pre-maximum wind 
just outside the shocked shell.

%

\section{High-energy emission from shocked matter}
\label{section_emission}


A strong shock may also provide high-energy photons. 
Recently the Fermi Large Area Telescope (Fermi/LAT) detected 
GeV gamma-ray emissions associated to classical novae 
\citep[e.g.,][]{abd10, ack14, che16js, li17mc}.
These gamma rays are considered to originate from
strong shocks \citep[see, e.g.,][for a recent review]{cho21ms}.

To explain GeV gamma-ray emissions from novae, two 
types of shock formation mechanisms have ever been proposed.
One is the collision between nova ejecta and circumstellar matter
(CSM; ejecta-CSM shock model or external shock model)
as observed in the symbiotic nova
V407 Cyg \citep[e.g.,][]{abd10} and the symbiotic recurrent nova 
RS Oph \citep[see][for the 2021 outburst]{che21cj, che21jm, acc22aa, hes22}.

The other type is for novae having a red dwarf (main-sequence star) companion. 
Such a close binary does not form massive circumstellar matter before 
a nova outburst. Thus, the hypothesis of ``internal shocks'' in nova ejecta
was proposed to explain hard X-ray emissions 
\citep[e.g.,][]{friedj87, llo92ob, muk01i}.
These authors assumed multiple shell ejecta that have different velocities.
If the inner shell (later ejected) has a larger velocity than that of
the outer shell (earlier ejected), the inner one can catch up with the 
outer one and form a shock wave. 
This idea is based on the assumption of multiple shell ejections. 

In this section, we apply our shock formation model to this
idea and examine whether it works or not
both for hard X-ray emission and GeV gamma-ray emission.


\begin{figure*}
\epsscale{0.85}
\plotone{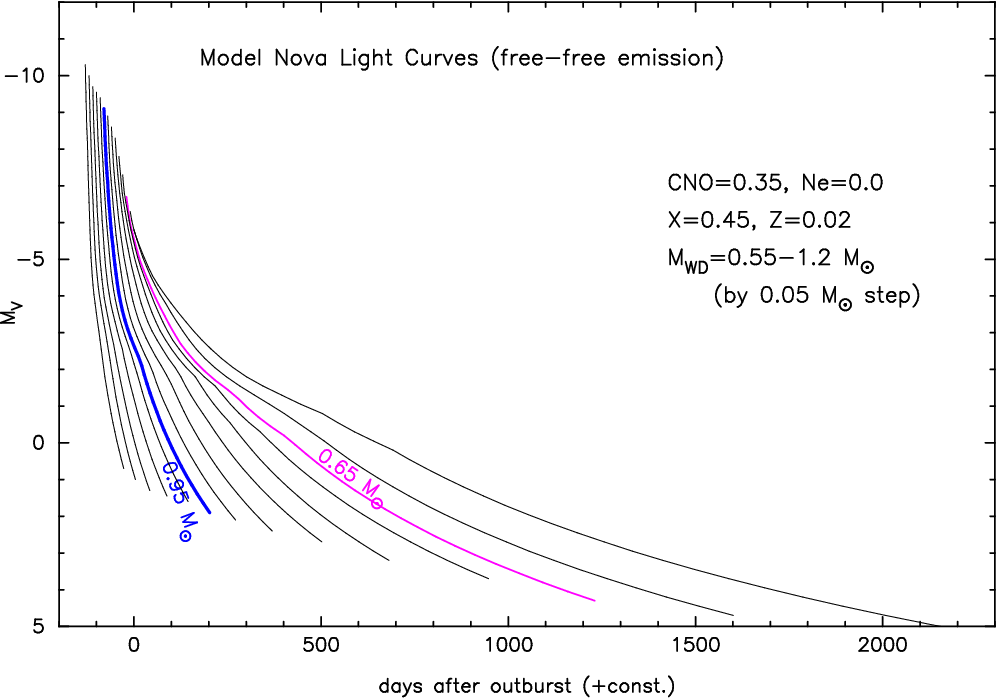}
\caption{
Absolute $V$ light curves of free-free emission
for various WD masses, i.e., 
$M_{\rm WD}= 0.55$, 0.6, 0.65, 0.7, 0.75, 0.8, 0.85,
0.9, 0.95, 1.0, 1.05, 1.1, 1.15, and $1.2~M_\sun$ with the chemical
composition of the hydrogen-rich envelope of $X=0.45$, $Y=0.18$, $Z=0.02$,
$X_{\rm C}=0.15$, and $X_{\rm O}=0.20$ \citep[see][]{hac06kb, hac16k}.
\citet{hac06kb} called this chemical composition ``CO nova 3 (CO3)''. 
The proportionality constant $A_{\rm ff}$ in equation
(\ref{free-free_flux_v-band}) was calibrated by
\citet{hac16k} and \citet{hac20skhs}.
Various nova parameters are calculated from the steady-state 
envelope solutions given by \citet{kat94h}.  The $0.95~M_\sun$ model
(blue line) corresponds to typical fast novae
while the $0.65~M_\sun$ model (magenta line)
is a representative for slow or very slow novae. 
\label{light_curve_combine_x45z02c15o20_absmag}}
\end{figure*}

\subsection{Hard X-ray Emission}
\label{section_hard_x-ray}

The temperature just behind the shock is estimated to be
\begin{eqnarray}
kT_{\rm sh}& \sim & {3 \over 16} \mu m_p 
\left( v_{\rm wind} - v_{\rm shock} \right)^2 \cr
& \approx & 1.0 {\rm ~keV~} 
\left( {{v_{\rm wind} - v_{\rm shock}} \over  
{1000 {\rm ~km~s}^{-1}}} \right)^2
\label{shock_kev_energy}
\end{eqnarray}
or
\begin{equation}
T_{\rm sh} \sim 1.1\times 10^7  {\rm ~K~}
\left( {{v_{\rm wind} - v_{\rm shock}} \over  
{1000 {\rm ~km~s}^{-1}}} \right)^2
\label{shock_temperature}
\end{equation}
where $k$ is the Boltzmann constant,
$T_{\rm sh}$ is the temperature just after the shock
\citep[see, e.g.,][]{met14hv}, 
$\mu$ is the mean molecular weight ($\mu =0.5$ for hydrogen plasma),
and $m_p$ is the proton mass.
Substituting our model values, we obtain the post-shock
temperature $k T_{\rm sh}$, as plotted in 
Figure \ref{shock_density_pp_interaction}b (blue line).

Mechanical energy of wind is converted to thermal energy
by the reverse shock at a rate of
\begin{eqnarray}
L_{\rm sh}& \sim & {{9}\over {32}} {\dot M}_{\rm wind} 
{{( v_{\rm wind} - v_{\rm shock} )^3} \over {v_{\rm wind}}} \cr
&=& 1.8\times 10^{37}{\rm ~erg~s}^{-1}
\left( {{{\dot M}_{\rm wind}} \over 
{10^{-4} ~M_\sun {\rm ~yr}^{-1}}} \right) \cr
 &  & \times
\left( {{{v_{\rm wind} - v_{\rm shock}} \over {1000{\rm ~km~s}^{-1}}}}
\right)^3
\left( {{{1000{\rm ~km~s}^{-1}} \over {v_{\rm wind}}} }\right), 
\label{shocked_energy_generation}
\end{eqnarray}
which was taken from \citet{met14hv}.
We also plot the time variation of 
$L_{\rm sh}$ by the red line 
in Figure \ref{shock_density_pp_interaction}b. 
Both the temperature $k T_{\rm sh}$ and luminosity $L_{\rm sh}$ quickly
increase at $t\sim 90$ days after outburst,
i.e., about 60 days after the optical peak. 

The converted thermal energy, $L_{\rm sh}$, quickly increases to 
$L_{\rm sh}\sim 10^{35}$erg s$^{-1}$ in 55 days 
while the thermal temperature is as low as 
$k T_{\rm sh}\sim 0.02$ keV at $t\sim 150$ day.
This low temperature comes from the small velocity difference, 
$v_{\rm wind}(t) - v_{\rm shock}(t)$, which is $\sim 140$ km s$^{-1}$
in equations (\ref{shock_kev_energy}) and (\ref{shock_temperature})
at $t\sim 150$ day.
The wind velocity $v_{\rm wind}(t)$ just in front
of the shock is calculated from 
$v_{\rm wind}(t)= v_{\rm ph}(t-t_{\rm ret})$ and
$t_{\rm ret}= (R_{\rm sh}(t) - R_{\rm ph}(t-t_{\rm ret}))
/v_{\rm ph}(t-t_{\rm ret})$, where 
$R_{\rm sh}(t)$ is the present position of shock,
$R_{\rm ph}(t-t_{\rm ret})$ and $v_{\rm ph}(t-t_{\rm ret})$
are the photospheric radius and velocity at $t-t_{\rm ret}$,
and $t_{\rm ret}$ is the retarded (look back) time, as can be seen in
Figure \ref{shock_density_pp_interaction}a.

  For temperatures of $\sim 0.01$--0.05 keV in Figure
\ref{shock_density_pp_interaction}b,
one may expect soft X-ray emissions from thermal plasma in nova ejecta.
Such soft X-rays are not observed because the hydrogen column density
$N_{\rm H}$ is still as high as $\sim 10^{23}$ cm$^{-2}$ at
$t\sim 90$ day as shown in Figure \ref{shock_density_pp_interaction}c.
Thus, our model for a moderately fast nova predicts no high-energy 
emissions from the shock because of a small difference in velocities
between the wind and shock, i.e., $v_{\rm wind}-v_{\rm shock}\ll 1000$
km s$^{-1}$.

Hard X-ray emissions are detected in some novae, such  
as in V838 Her \citep{llo92ob}, in V1974 Cyg \citep{bal98ko},
and in V382 Vel \citep{muk01i} and interpreted as the shocked ejecta 
emission. 

\subsubsection{Velocity Systems in V1974 Cyg}
\label{comparison_various_feature_v1974_cyg}

First, we make a qualitative comparison of our model velocities
with those in the fast nova V1974 Cyg,
of which the three velocity components were well observed, 
i.e., $v_{\rm early}\sim 5000$ km s$^{-1}$,
$v_{\rm shock}\sim 800$ km s$^{-1}$ (principal spectra), 
and $v_{\rm wind}\sim 1500$ km s$^{-1}$ (diffuse-enhanced spectra)
\citep[e.g.,][]{cho97gp, cas04lr}. 
The velocity of the principal system increases from 800 km s$^{-1}$ 
(post-maximum) to 1700 km s$^{-1}$ (at $\sim 50$ day).
Correspondingly, the velocity of the diffuse-enhanced system 
goes up from 1500 km s$^{-1}$ to 2900 km s$^{-1}$.
This increasing trend is similar to that of our shocked ejecta model. 

\citet{cho97gp} discussed the empirical $v_{\rm p}$-$t_3$ relation,
i.e., $t_3=42$ day for V1974 Cyg and $v_{\rm p}\approx 800$ km s$^{-1}$
which is derived from equation (\ref{vp_vs_t3}). 
They discussed that the assignment of $v_{\rm p}\sim 800$ km s$^{-1}$
at post-maximum is reasonable as a principal system.
More rapidly declining novae have larger velocities
of the principal systems.  

Simply substituting 
$v_{\rm wind} - v_{\rm shock} =v_{\rm d} - v_{\rm p} =
1500$ -- 800 = 700 km s$^{-1}$ or $=2900$ -- 1750 = 1150 km s$^{-1}$
into equation (\ref{shock_kev_energy}), we obtain $k T_{\rm sh}
\sim 0.5$ -- 1.3 keV, which is broadly consistent with
the temperature estimated by \citet{bal98ko}.
The converted thermal energy generation rate is calculated to be
as large as $L_{\rm sh}\sim 1.8 \times 10^{37} \times 1.0 
\times (1.15)^3/2.9 \sim 1 \times 10^{37}$ erg s$^{-1}$
for $\dot{M}_{\rm wind}\sim 1\times 10^{-4} ~M_\sun$ yr$^{-1}$. 

Thus, we can reproduce the temperature of k$T_{\rm sh}\sim 1$--2 keV
if we take the observed velocities of $v_{\rm d} - v_{\rm p}$ 
instead of our model value of $v_{\rm wind} - v_{\rm shock}$. 
Our original shock formation model based on \citet{kat22sha}'s
calculation has problems in explaining hard X-ray emission of V1974 Cyg.

If we roughly assume that $v_{\rm d}\sim 2 v_{\rm p}$ as summarized by
\citet{mcl42}, we have $k T_{\rm sh}\sim 1 {\rm ~keV} (v_{\rm p}/
{\rm 1000 ~km~s}^{-1})^2$.  We further assume that hard X-ray
emission is detected for $k T_{\rm sh}\gtrsim 0.5$ keV. Then, we
constrain the nova speed class from equation
(\ref{vp_vs_t3}), that is, $t_3 \lesssim 50$ day,
under the condition of $v_{\rm p} \gtrsim 700$ km s$^{-1}$.
The $t_3 \lesssim 50$ day speed class ranges from
very fast, 
fast, 
moderately fast novae.  
We expect the detection of hard X-ray emission from 
novae to have a speed class of $t_3 \lesssim 50$ day.
V838 Her ($t_3=4$ day), V1974 Cyg (43 day), and V382 Vel (12 day)
all satisfy the $t_3 \lesssim 50$ day speed class
\citep[$t_3$ of these novae are taken from][]{schaefer18}.





\begin{figure}
\epsscale{1.15}
\plotone{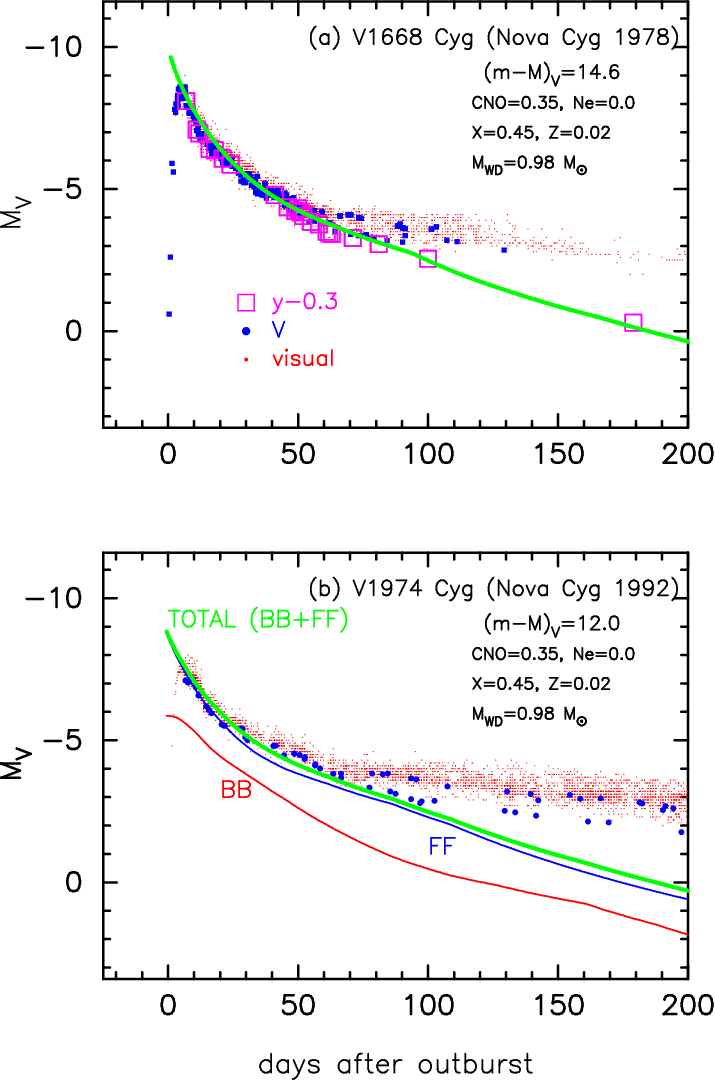}
\caption{
Same as those in Figure \ref{opt_gamma_ray_abs_magnitudes}, but for
a $0.98~M_\sun$ WD with CNO enhancement (CO3).
Only the decay phase (after the optical maximum) of the steady-state model 
in Figure \ref{light_curve_combine_x45z02c15o20_absmag}
\citep{kat94h, hac06kb} is plotted.
The solid red line denotes blackbody emission from the photosphere
(labeled BB).  The solid blue line is free-free emission (labeled FF).
The thick solid green lines are the total flux (labeled TOTAL) 
of BB and FF.    
(a) The $0.98~M_\sun$ WD model light curve is fitted with the visual
(red dots), $V$ (filled blue circles), and $y$ (open magenta squares)
light curves of V1668 Cyg. 
The visual magnitudes are taken from the archive of the American
Association of Variable Star Observers (AAVSO),
and those of $V$ magnitudes are from \citet{dip81}, \citet{pic84},
\citet{hop79}, \citet{kol80}, \citet{mal79}, and \citet{due80}.
The data of the $y$ magnitudes are from \citet{gal80}.
(b) Same as panel (a) but for V1974 Cyg.
The visual data are taken from AAVSO, the $V$ data are taken from
\citet{cho93hu}.
\label{opt_gamma_ray_abs_magnitudes_v1668_cyg}}
\end{figure}

\subsection{G{\lowercase{e}}V Gamma-Ray Emission from a Shock}
\label{gev_gamma_ray_emission}

The shock properties that produce GeV gamma-ray emission constrain
\begin{equation}
L_\gamma = \epsilon_{\rm nth} \epsilon_\gamma L_{\rm sh} \sim 10^{35} -
10^{36} {\rm ~erg~s}^{-1},
\label{observed_gamma_ray_flux}
\end{equation}
from observation \citep{ack14}, where $\epsilon_{\rm nth} \lesssim 0.1$
is the fraction of the shocked thermal energy used to accelerate 
relativistic nonthermal particles, $\epsilon_\gamma \lesssim 0.1$
the fraction of this energy radiated in the Fermi/LAT band
\citep[typically $\epsilon_{\rm nth} \epsilon_\gamma 
\lesssim 0.03$;][]{met15fv}.  This simply requires
\begin{equation}
L_{\rm sh} \sim 10^{37} - 10^{38} {\rm ~erg~s}^{-1}.
\label{shocked_generation_energy_flux}
\end{equation}

We may not expect such a large flux of GeV gamma-ray emission
from the shocked ejecta in \citet{kat22sha}'s model
because of its small velocity difference,
$v_{\rm wind} - v_{\rm shock} \lesssim 300$ km s$^{-1}$ or, in other words,
its low shocked luminosity of 
$L_{\rm sh}\sim 10^{35}$ erg s$^{-1}$ in Figure 
\ref{shock_density_pp_interaction}b.

Here, we examine Fermi acceleration as the source of GeV gamma rays
\citep[e.g.,][]{abd10}.  \citet{li17mc} examined the two processes
proposed for gamma rays from novae and preferred proton collisions
(hadronic scenario) for V5856 Sgr, rather than inverse 
Compton/Bremsstrahlung emission (leptonic scenario).  

We estimate the timescale of proton-proton ($pp$)
collisions a day after the optical maximum. 
At radius $R\sim 5\times 10^{12}$ cm $=71~R_\sun$, 
it is about $t_{pp}\approx 1/[4 n(R) c \sigma_{pp}]\sim 1000$ s, 
where $c$ is the speed of light, and
$\sigma_{pp}=3\times 10^{-26}$ cm$^2$ is the proton-proton cross section
\citep{abd10}. 
The number density $n(R)\sim 2.6\times 10^{11}$ cm$^{-3}$
is calculated from the wind mass-loss rate of our model, 
${\dot M}_{\rm wind}
= 4 \pi r^2 \rho v_{\rm ph} \sim 1.3 \times 10^{-4}~M_\sun$ yr$^{-1}$. 

Eight days later, the shock expands to $R\sim 2.4\times 10^{13}$ cm. 
The timescale of proton-proton collision becomes longer but still as 
short as $t_{pp} \sim 10,000$ s $\sim 0.11$ day.
About a month later, the shock expands to $R\sim 7\times 10^{13}$ cm
$=1000~R_\sun$ and the timescale becomes $t_{pp} \sim 86,000$ s $\sim 1$ day.
The collision timescale is short enough for all the protons 
to interact with each other and to produce $\pi^0$ in a day. 
The temporal variation of $n(R)$ is plotted
in Figure \ref{shock_density_pp_interaction}c. 
If we constrain $t_{pp} < 1$ day for gamma-ray detection,
its duration is about 35 days.  Thus, the shock satisfies the timescale
of $t_{pp}$.  However, the shocked energy is too small to explain
the GeV gamma-ray fluxes.

\citet{li17mc} estimated the gamma-ray flux from the nova 
ASASSN-16ma (V5856 Sgr) to be $F_\gamma= (4-8)\times 10^{-10}$
erg cm$^{-2}$ s$^{-1}$.  With the assumed distance to the nova of 4.2 kpc,
they obtained the total flux of gamma rays $L_\gamma \sim
(0.8-1.6)\times 10^{36}$ erg s$^{-1}$.  This corresponds to about a half
or fourth of the kinetic energy of wind ($L_{\rm kin}\sim 3\times 10^{36}$
erg s$^{-1}$) in our model (see Figure \ref{mdot_radius_velocity}b).
Even if all the kinetic energy $L_{\rm kin}$ is converted to the shocked
energy $L_{\rm sh}$, the gamma-ray flux $L_{\gamma}$ is too small
to be compatible with the observation because the conversion rate
is as small as $L_\gamma / L_{\rm sh} \sim 0.03$.

\subsubsection{Velocity Systems in V5856 Sgr}
\label{velocity_system_v5856_sgr}

For V1974 Cyg in section \ref{comparison_various_feature_v1974_cyg}
(and V382 Vel in section \ref{velocity_systems_v382_vel}),
we adopted $v_{\rm p}$ and $v_{\rm d}$
from each author's definition.  However, no clear definitions of
$v_{\rm p}$ and $v_{\rm d}$ are given in V5856 Sgr.

\citet{ayd20ci} discussed the relation between old McLaughlin's
definition on photographic plates and modern one-dimensional spectra
extracted from CCDs.  We basically followed their descriptions. 
\citet{ayd20ci} wrote ``all of the novae show a similar spectral evolution: 
before the optical peak, the emission lines show P Cygni profiles
with absorption troughs at velocities $\sim$ 200--1000 km s$^{-1}$.
After the optical peak, a broad emission component
emerges with the velocities $> 1000$ km s$^{-1}$ (more than twice
the velocity of the pre-maximum component).''
They called the pre-maximum P Cygni profile the slow component
and the post-maximum broad emission + higher-velocity P Cygni absorption
the fast component.  They concluded that these are the same as
the pre-maximum and diffuse-enhanced systems of \citet{mcl44}.

\citet{ayd20ci} also pointed out, on the spectra of V906 Car at the maximum,
the coexistence of a new component at a velocity of around
$300$ km s$^{-1}$ and the slower pre-maximum component at $200$ km s$^{-1}$.
They concluded that this intermediate-velocity ($300$ km s$^{-1}$)
system is what McLaughlin calls the principal system \citep{mcl44, pay57}.
They called this the intermediate component.

The slow component is decelerating (slowing down) in the pre-maximum phase,
but it appears to be accelerating after the optical maximum
\citep[see, e.g., Figure 11 of ][]{ayd20ci}.
This trend was interpreted as the replacement of the slow component 
by the intermediate component \citep[e.g.,][]{ayd20ci}.  In our model, 
the shocked shell (McLaughlin's principal shell), which
forms a narrow and deep P Cygni profile, is accelerated by the collision of
the fast wind \citep[see also][]{friedj87}.  Therefore, the accelerated 
slow component or intermediate component possibly corresponds 
to the principal system.

\citet{li17mc} reported P Cygni absorption profiles on H$\alpha$ line
of V5856 Sgr,
from which we derived the velocity of principal system post-maximum.
We measured the velocity at the deepest absorption trough in their
Supplementary Figure 2, which is roughly 
 $-600$ km s$^{-1}$ at $-8$ day, ($=$ 8 day before the optical maximum, 
discovery date $=$ 25 October 2016= $-14$ day),
 $-500$ km s$^{-1}$ ($-6$ day),
 $-800$ km s$^{-1}$ ($+2$ day),
 $-800$ km s$^{-1}$ ($+4$ day).
We assign this velocity to $v_{\rm p}= -800$ km s$^{-1}$ post-maximum
($+4$ day).  They also reported that the higher-velocity component
appeared and reached 2200 km s$^{-1}$, so we adopted 
$v_{\rm d}= 2200$ km s$^{-1}$ post-maximum ($+4$ day).
The velocity difference is $v_{\rm d}-v_{\rm p}= 2200 -800 = 1400$ km s$^{-1}$
post-maximum ($+4$ days) of V5856 Sgr.

Then, the shocked thermal temperature
is k$T_{\rm sh}= 2.0$ keV from equation (\ref{shock_kev_energy}),
which is high enough to emit hard ($\gtrsim 1$ keV) X-rays.
The shocked thermal energy generation rate is estimated to be
$L_{\rm sh}= 1.8\times 10^{37} (1.4)^3/2.2 \times 1.0 = 2.2\times 10^{37}$
erg s$^{-1}$ from equation (\ref{shocked_energy_generation})
for $\dot {M}_{\rm wind}= 1.0\times 10^{-4} ~M_\sun$ yr$^{-1}$.
GeV gamma rays were detected from $+1$ day to $+14$ day.
This shocked thermal luminosity gives a ratio of 
$L_\gamma / L_{\rm sh} \sim 0.04$. 

To summarize, in the V5856 Sgr observation, the velocity difference
is as large as $v_{\rm wind} - v_{\rm shock}
= v_{\rm d} - v_{\rm p} \sim 1400$ km s$^{-1}$ and, as a result,
the shock luminosity is large enough to reproduce 
$L_\gamma \sim 0.04 L_{\rm sh} \sim 1 \times 10^{36}$ erg s$^{-1}$
for the wind mass-loss rate of $\dot{M}_{\rm wind}=1.0\times 10^{-4}
~M_\sun$ yr$^{-1}$.  In what follows, we reproduce the gamma-ray 
luminosity assuming $v_{\rm shock}= v_{\rm p}$ and
$v_{\rm wind}= v_{\rm d}$, which are measured from optical spectra of novae.

\section{Discussion}
\label{discussion}


\subsection{Super-Eddington Luminosity}
\label{super_eddington}

A number of nova evolution calculations have been presented, 
but none of them explained the observed brightnesses of novae that 
often exceed the Eddington limit \citep[e.g.,][]{del20i}.  
In addition, nova spectra sometimes show a flat pattern, $F_\nu\sim$ constant 
against the frequency $\nu$, different from the blackbody spectrum 
\citep[e.g.,][ for one of the brightest novae V1500 Cyg: 
soon after the optical maximum]{gal76, enn77}.
\citet{hac06kb} pointed out that free-free emission 
from nova ejecta outside the photosphere dominates the nova
spectra that mainly contribute to the optical luminosity.
They proposed a description formula of the free-free flux 
from nova ejecta as
\begin{equation}
F_\nu \propto {{\dot M^2_{\rm wind}} 
\over{v^2_{\rm ph} R_{\rm ph}}},
\label{free-free_flux_definition}
\end{equation}
and define the $V$-band flux $L_{V, \rm ff}$ as 
\begin{equation}
L_{V, \rm ff} = A_{\rm ff} ~{{\dot M^2_{\rm wind}} 
\over{v^2_{\rm ph} R_{\rm ph}}},
\label{free-free_flux_v-band}
\end{equation}
where the coefficient $A_{\rm ff}$ was determined by 
\citet{hac10k, hac15k, hac16k} and \citet{hac20skhs}.
The physical meaning of this formulation is described in more detail 
in \citet{hac20skhs}. 
Then, the total $V$-band flux is 
\begin{equation}
L_{V, \rm total} = L_{V, \rm ff} + L_{V, \rm ph},
\label{luminosity_summation_flux_v-band}
\end{equation}
where $L_{V, \rm ph}$ is the $V$-band flux of $L_{\rm ph}$. 

The light curve analysis by Hachisu and Kato, based on 
the continuous mass loss,  
reproduces a number of observed multiwavelength light 
curves of novae, including the super-Eddington phase, 
which strongly indicates 
that the current theoretical understanding of mass loss is reasonable
\citep[e.g.,][]{hac16k, hac18kb}. 

We plot the absolute $V$ magnitudes of these fluxes for
\citet{kat22sha}'s model in Figure
\ref{opt_gamma_ray_abs_magnitudes}.
Here, $m_{V, \rm total}$, $m_{V, \rm ff}$, and $m_{V, \rm ph}$
are the absolute magnitudes obtained from 
$L_{V, \rm total}$, $L_{V, \rm ff}$,
and $L_{V, \rm ph}$, respectively.

It is interesting that three light curves of $m_{V, \rm total}$, 
$m_{V, \rm ff}$, and $m_{V, \rm ph}$ have a very similar 
shape. The photospheric $V$
magnitude does not exceed the Eddington limit ($M_V \sim -6$), but
the free-free $V$ ($M_V \sim -7.2$) and total $V$ ($M_V \sim -7.5$)
magnitudes are much brighter than the Eddington limit ($M_V \sim -6$),
where $M_V$ is the absolute $V$ magnitude.
%


\begin{figure}
\gridline{\fig{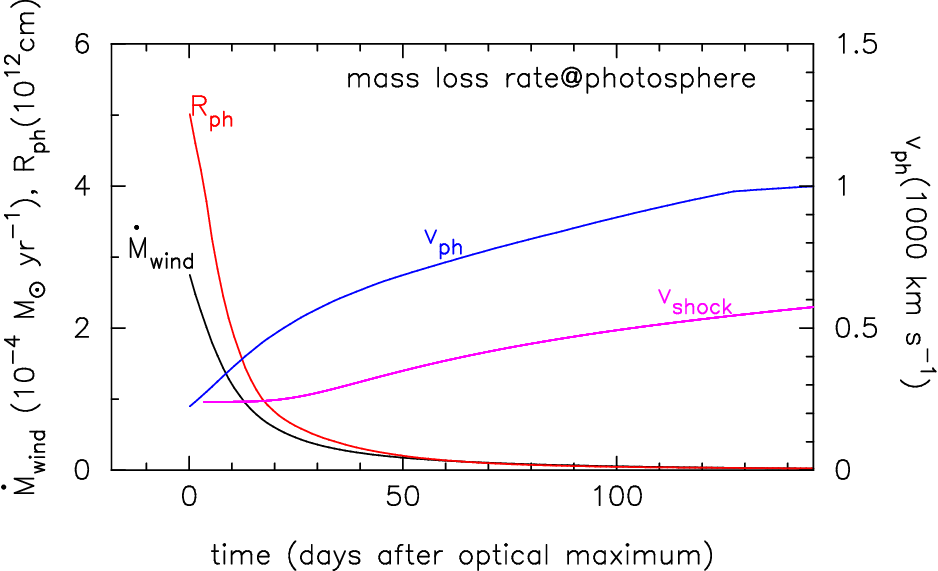}{0.47\textwidth}{(a)}
          }
\gridline{
          \fig{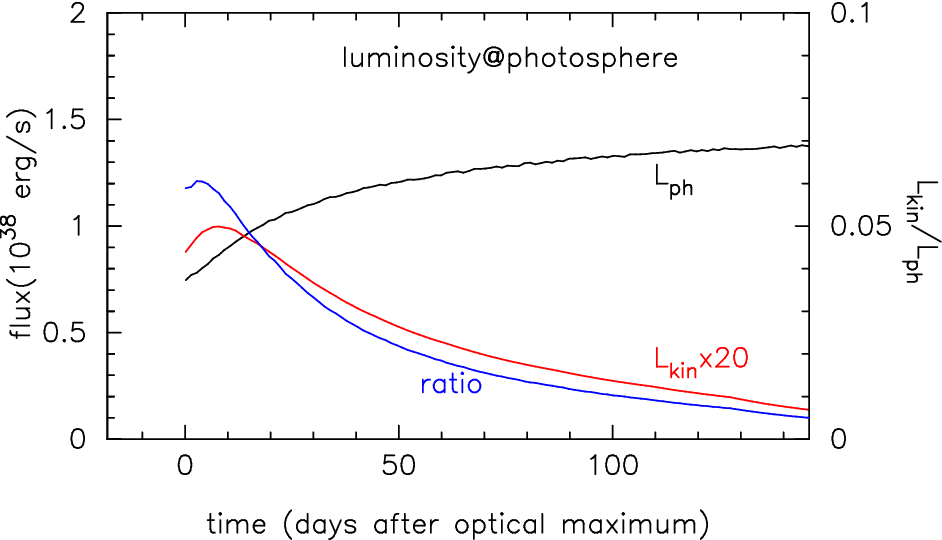}{0.47\textwidth}{(b)}
          }
\caption{
Same as those in Figure \ref{mdot_radius_velocity}, but for 
a $0.98~M_\sun$ WD with CNO enhancement (CO3).
The time $t=0$ correspond to the optical maximum. 
Only the decay phase (after the optical maximum) of steady-state envelope
model is plotted.
\label{mdot_radius_velocity_v1668_cyg}}
\end{figure}


\begin{figure}
\gridline{\fig{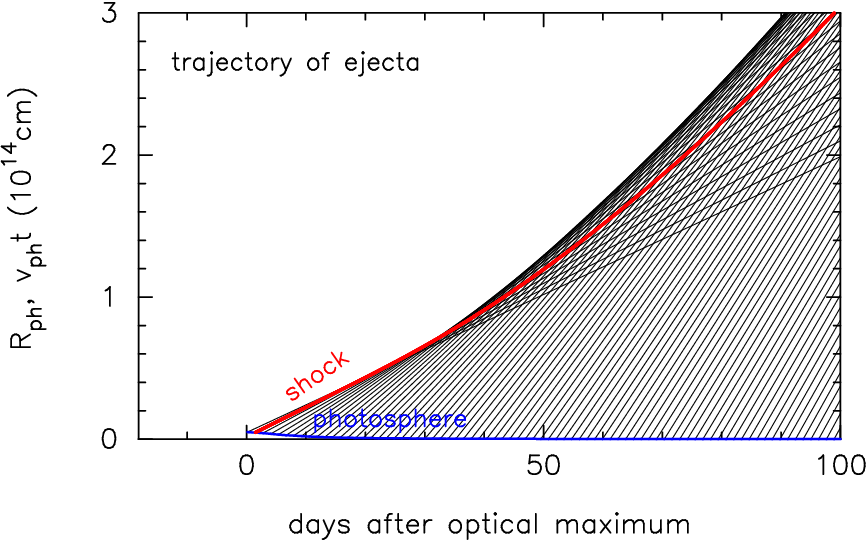}{0.47\textwidth}{(a)}
          }
\gridline{
          \fig{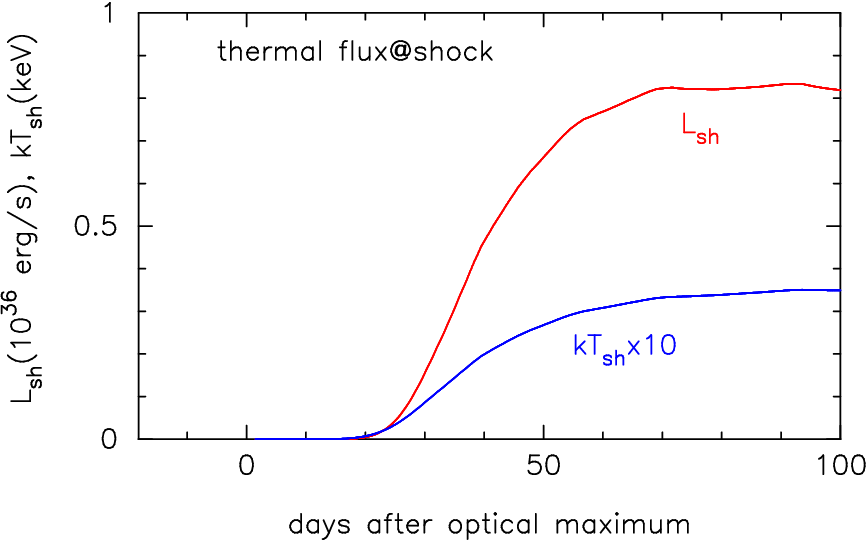}{0.47\textwidth}{(b)}
          }
\gridline{
          \fig{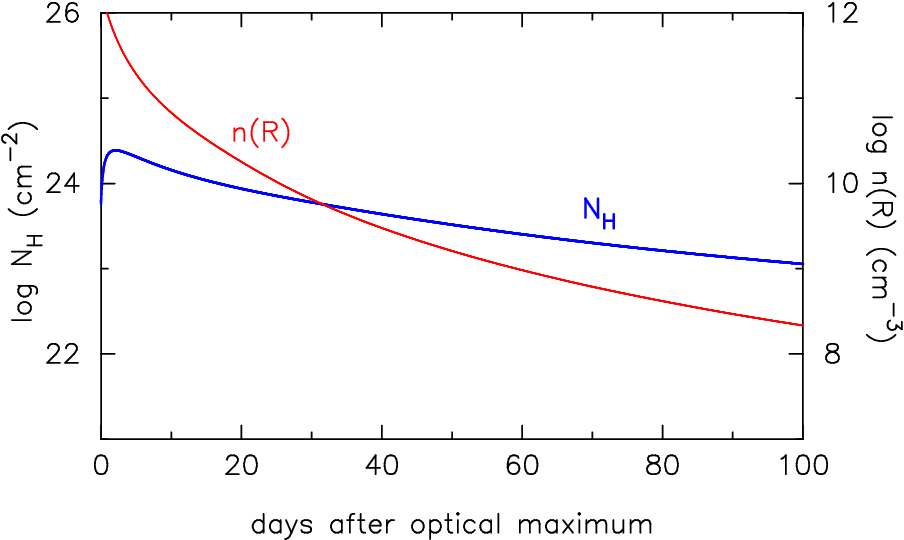}{0.47\textwidth}{(c)}
          }
\caption{Same as in Figure \ref{shock_density_pp_interaction},
but for a $0.98~M_\sun$ WD with CNO enhancement (CO3).
Only the decay phase (after the optical maximum) of steady-state envelope
model is plotted.
{\bf (a)} A strong shock forms soon after the optical maximum
and expands outward at a speed of 250--600 km s$^{-1}$.
{\bf (b)} Both the shocked luminosity $L_{\rm sh}$ (red)
and temperature $k T_{\rm sh}$ (blue)
quickly rise up $\sim 30$ days after the optical maximum.
{\bf (c)} Temporal changes in the column density $N_{\rm H}$ and
number density $n(R)$, slightly denser 
than those of the 1.0 $M_\sun$
model in Figure \ref{shock_density_pp_interaction}c.
\label{density_pp_interaction_v1668_cyg}}
\end{figure}

\subsection{Enhancement of CNO Elements}
\label{enhancement_cno_elements}

Heavy-element enhancement such as C, N, O, and Ne was frequently
observed in ejecta of classical novae \citep[e.g.,][]{geh98}. 
On the other hand, \citet{kat22sha}'s model does not show heavy-element
enrichment because elemental mixing with WD core material
does not occur in 1D calculations, and they did not include
the additional dredging-up mechanism. 
In this subsection, we examine how such heavy-element enhancement 
alters shock wave and gamma-ray emission. 

Here we use the steady-state wind model calculated by \citet{kat94h}
because no fully self-consistent models are presented for 
the case of heavy-element enrichment. 
This steady-state model is a good approximation of 
the decay phase of a nova outburst \citep{kat22sha}. 

Figure \ref{light_curve_combine_x45z02c15o20_absmag} shows
optical $V$ light curves of novae (free-free emission)
but for a CNO enhancement
model, $X=0.45$, $Y=0.18$, $Z=0.02$, $X_{\rm C}=0.15$,
and $X_{\rm O}=0.20$ by mass weight
\citep[see][for detail]{hac06kb, hac10k, hac16k, hac17k}. 
They dubbed this chemical composition model ``CO nova 3 (CO3).''
For comparison, we choose the $0.98~M_\sun$ WD model, which corresponds to
a typical fast nova and that reproduces the multiwavelength light curves
of V1668 Cyg \citep[e.g.,][]{hac16k} as shown in Figure
\ref{opt_gamma_ray_abs_magnitudes_v1668_cyg}a. 
The temporal variations of photospheric values 
in Figures \ref{mdot_radius_velocity_v1668_cyg}
and \ref{density_pp_interaction_v1668_cyg} are
essentially similar to those in Figures \ref{mdot_radius_velocity}
and \ref{shock_density_pp_interaction}, respectively.
We plot physical values after the optical maximum,
because a shock is generated only after the optical maximum,
and a steady state is established after the optical maximum \citep{kat22sha}.

The largest difference is the duration of the wind phase, 
about 3 times shorter, 
in the 0.98 $M_\sun$ (CO3) model than in our model in section
\ref{section_model}. 
The photospheric velocity is about $\sim 200$ km s$^{-1}$ ($\sim 30$\%)
faster.  The wind mass-loss rate is slightly larger.
These differences mainly come from (i) less fuel of hydrogen
($X=0.65$ versus $X=0.45$), (ii) smaller ignition mass 
in heavy-element enrichment models \citep[e.g.,][]{chen19},
and (iii) larger acceleration owing to the radiation-pressure gradient. 
We have already discussed these differences in more detail 
in \citet{kat22sha}.

The velocity difference is still as low as 
$v_{\rm wind} - v_{\rm shock}
= v_{\rm ph}(t-t_{\rm ret}) - v_{\rm sh}(t)
 \approx 452 - 311 = 141$ km s$^{-1}$ 
at $t=40$ day as can be seen in Figures
\ref{mdot_radius_velocity_v1668_cyg}a and
\ref{density_pp_interaction_v1668_cyg}a, so that 
the temperature is as low as $k T_{\rm sh}\sim 0.02$ keV,
essentially the same as in Figure \ref{shock_density_pp_interaction}b.
We expect soft X-ray emission from the shocked matter.
Such soft X-rays are not observable due to heavy absorption 
($\log N_{\rm H} \sim 23$--24) as shown in Figure
\ref{density_pp_interaction_v1668_cyg}c.

As for GeV gamma-ray emission,
the shocked thermal energy generation rate is as small as
$L_{\rm sh}\lesssim 0.01\times 10^{36}$ erg s$^{-1}$ at post-maximum
(Figure \ref{density_pp_interaction_v1668_cyg}b), much
smaller than $L_\gamma \sim (0.8-1.6)\times 10^{36}$ erg s$^{-1}$
\citep[for V5856 Sgr,][]{li17mc}.
Even if we use a CNO-enriched model of $0.98 ~M_\sun$ WD,
we cannot reproduce the early GeV gamma-ray emission like in V5856 Sgr.

\subsection{V1974 Cyg 1992}
\label{v1974_cyg_1992}

\subsubsection{Model Light Curve Fitting of V1974 Cyg}
\label{model_fitting_v1974_cyg_wd_mass}

Figure \ref{opt_gamma_ray_abs_magnitudes_v1668_cyg}b shows the absolute
magnitudes of $V$ light curves for each component, i.e., the photospheric
(red line, labeled BB), free-free emission (blue, FF),
and the total of them (green, TOTAL).
This model light curve (green line) reproduces well the observed
$V$ light curve (filled blue circles) of V1974 Cyg if we adopt
the $V$-band distance modulus, $(m-M)_V= A_V + 5 \log (d/10{\rm ~pc})
= 0.93 + 5 \log (1.6{\rm ~kpc}/10{\rm ~pc}) = 12.0$,
where, the $V$-band absorption of $A_V=0.93$ is taken
from \citet{hac19ka} and \citet{hac21k}
and the distance of $d=1.6$ kpc is from Gaia early data release 3
\citep[EDR3;][]{bai21rf}. 

Because the $V$-band distance modulus is fixed to be $(m-M)_V= 12.0$,
we have only one degree of freedom, that is, shifting the model light curves
back and forth horizontally. In the figure, we fit our total $V$
model light curves with the optical maximum.  The model light curve
labeled TOTAL reasonably follows the $V$ data of V1974 Cyg
in an early phase (up to $\sim 90$ days after the optical maximum).
After that, the model light curve deviates downward 
largely from the visual and $V$ data.
This is because strong emission lines such as [\ion{O}{3}]
significantly contribute to the $V$-band flux as discussed by \citet{hac06kb}.
If we could use the intermediate $y$-band magnitude, which avoids 
strong emission lines such as [\ion{O}{3}], our model light curve 
follows well the continuum flux of the nova
even in the nebular phase, as seen in Figure
\ref{opt_gamma_ray_abs_magnitudes_v1668_cyg}a of V1668 Cyg.
Thus, the continuum flux of the nova decreases almost along with
the model light curve.
Therefore, we adopt the 0.98 $M_\sun$ as a good model both for V1668 Cyg
and V1974 Cyg.

We should note that V1974 Cyg was identified as a neon nova
\citep[e.g.,][for the neon abundance]{sho97sa, van05ss}.  In Figure
\ref{opt_gamma_ray_abs_magnitudes_v1668_cyg}b, we used a CNO-rich 
composition (from our template models in Figure
\ref{light_curve_combine_x45z02c15o20_absmag})
instead of a neon-rich composition because the 
enrichment of neon with unchanged hydrogen and CNO mass fractions
affects the nova light curves very little in our model light curves.
This can be justified by the result that neon is not relevant to
either nuclear burning (the CNO cycle)
or the opacity \citep[see][for more detail]{hac16k}.

\subsubsection{Absorption/Emission Line Systems in V1974 Cyg}
\label{absorption_line_system_v1974_cyg}

Our 0.98 $M_\sun$ WD model reasonably reproduced the $V$ 
(as well as UV 1455\AA\  and soft X-ray) light curve
\citep[see][for more details]{hac16k}.  
However, the velocities of absorption/emission line systems
are largely different from the observation.
We summarize the model versus observation as follows.
Principal system: 240 versus 800 km s$^{-1}$ just at post-maximum,
400 versus 1750 km s$^{-1}$ at $\sim 60$ days after maximum.
Diffuse enhanced system: 240 versus 1500 km s$^{-1}$ just at post-maximum,
580 versus 2900 km s$^{-1}$ at $\sim 60$ days after maximum.
The observed velocities, which are taken from \citet{cho97gp} and
\citet{cas04lr}, are 4 times larger than those in our model.

If our interpretation of $v_{\rm shock}= v_{\rm p}$ and
$v_{\rm wind}= v_{\rm d}$ is correct, our steady-state wind model
lacks some additional acceleration mechanisms inside or outside the 
photosphere.\footnote{
Note that the revised opacity tables (OPAL or OP in the 1990s) drastically 
changed nova calculations, and the obtained expansion 
velocities are multiplied.} 
In what follows, we adopt $v_{\rm shock}= v_{\rm p}$ and 
$v_{\rm wind}= v_{\rm d}$ and discuss the energy and temperature
of high-energy emissions.

\subsubsection{Hard X-Ray Emission in V1974 Cyg}
\label{hard_x_ray_emission_v1974_cyg}

As already discussed in section \ref{section_hard_x-ray},
we obtain sufficiently high shocked temperatures of
$k T_{\rm sh}\sim 1$ -- 2 keV, which is
consistent with \citet{bal98ko}'s estimate, if we use 
the observed principal ($v_{\rm p}$) and diffuse-enhanced
($v_{\rm d}$) velocities in V1974 Cyg.   
Then, the thermal energy generation rate is as large as
$L_{\rm sh}\sim 1.5 \times 10^{37}$ erg s$^{-1}$ just at post-maximum
for the wind mass-loss rate of $\dot{M}_{\rm wind}= 1.6\times
10^{-4} ~M_\sun$ yr$^{-1}$.  It decreases to
$L_{\rm sh}\sim 2.0 \times 10^{36}$ erg s$^{-1}$ 60 days after the maximum
for the wind mass-loss rate of $\dot{M}_{\rm wind}= 0.12\times
10^{-4} ~M_\sun$ yr$^{-1}$, which is large
enough to explain the hard X-ray flux of
$L_{\rm hx}\sim (0.8-2)\times 10^{34}$ erg s$^{-1}$
estimated by \citet{bal98ko}.  The conversion rate to X-rays in our model
is as small as $L_X / L_{\rm sh}\sim 0.01$ at day $+60$.

The column density of hydrogen is estimated from 
$M_{\rm shell}= 4 \pi R_{\rm sh}^2 \rho h_{\rm shell}$,
where 
$\rho$ is the density
in the shocked shell, and $h_{\rm shell}$ the thickness of the shocked shell.
If we take an averaged velocity of shell $v_{\rm sh}=
v_{\rm shell}= v_{\rm shock}$,
the shock radius is calculated from $R_{\rm sh}(t)= v_{\rm shock}\times t$.
This reads
\begin{eqnarray}
N_{\rm H} & = & {{X \over m_p} {{ M_{\rm shell} } 
\over {4 \pi R^2_{\rm sh}}}} \cr
 & \approx & 4.8\times 10^{22} {\rm ~cm}^{-2} 
\left({X \over {0.5}}\right)
\left( {{M_{\rm shell}} \over {10^{-5} M_\sun}} \right) 
\left( {{R_{\rm sh}} \over {10^{14} {\rm ~cm}}} \right)^{-2}
\cr
 & \approx & 6.4 \times 10^{20} {\rm ~cm}^{-2} 
\left({X \over {0.5}}\right)
\left( {{M_{\rm shell}} \over {10^{-5} M_\sun}} \right) \cr
& & \times 
\left( {{v_{\rm shell}} \over {1000 {\rm ~km~s}^{-1}}} \right)^{-2}
\left( {{t} \over {100~{\rm day}}} \right)^{-2}.
\label{column_density_hydrogen_time}
\end{eqnarray}
We assume that, when the column density becomes
as low as $N_{\rm H}\sim 10^{21}$ cm$^{-2}$,
hard X-rays ($1-10$ keV) can diffuse out from the shocked shell.
Then, we have the emergence time of 
\begin{eqnarray}
t & \approx & 80 {\rm ~day} \left({X \over {0.5}}\right)^{-{1 \over 2}}
\left( {{M_{\rm shell}} \over {10^{-5} M_\sun}} \right)^{{1 \over 2}} \cr
& & \times 
\left( {{N_{\rm H}} \over {10^{21} ~{\rm cm}^{-2}}} \right)^{-{1 \over 2}}
\left( {{v_{\rm shock}} \over {1000 ~{\rm km} ~ {\rm s}^{-1}}} \right)^{-1}.
\label{hard_xray_emerging_time}
\end{eqnarray}
We obtain the ejecta mass of $M_{\rm ej}=0.95\times 10^{-5} ~M_\sun$
from the light curve fitting in Figure
\ref{opt_gamma_ray_abs_magnitudes_v1668_cyg}b.
The average shock velocity is $v_{\rm shock}= v_{\rm p}=(800 + 1750)/2
\sim 1300$ km s$^{-1}$ from the principal absorption system in V1974 Cyg
\citep{cho97gp}.  Substituting these two values into equation
(\ref{hard_xray_emerging_time}), we have $t\sim 63$ day
for $N_{\rm H}\sim 10^{21}$ cm$^{-2}$.
This is consistent with the hard X-ray emergence time
$t \sim 60$ day of V1974 Cyg obtained by \citet{bal98ko}. 

Just at post-maximum ($t\sim 1$ day),
we obtain $M_{\rm shell}\approx 0.05 \times 10^{-5} ~M_\sun$
and $R_{\rm sh}\sim 1\times 10^{13}$ cm from the fitted model.
Then, this gives $N_{\rm H}= 2\times 10^{23}$ cm$^{-2}$ for $X= 0.45$.
Therefore, we could not detect hard X-ray emission near the optical 
maximum due to absorption by hydrogen behind the shock.

The optical depth for X-ray is estimated to be
\begin{eqnarray}
\tau_X & \approx & 8\times 10^3 \left( {{{\dot M}_{\rm wind}} \over
{10^{-4} M_\sun ~ {\rm yr}^{-1}}} \right)
\left( {{r} \over {10^{13} ~{\rm cm}}} \right)^{-1} \cr
& & \times 
\left( {{v_{\rm wind}} \over {400 ~{\rm km} ~ {\rm s}^{-1}}} \right)^{-1}
\left( {{E_X} \over {\rm keV}} \right)^{-2},
\label{hard_xray_optical_depth}
\end{eqnarray}
which is taken from equation (9) of \citet{li17mc}.
The optical depth is as high as 
$\tau_{\rm X}\sim 10^3( E_X/ {\rm keV} )^{-2}$
at post-maximum,
but it decreases to $\tau_{\rm X}\sim 10 ( E_X/ {\rm keV} )^{-2}$
at $\sim 60$ days after maximum.
Thus, hard X-ray emission could be observable at $\sim 60$ days
after the maximum and later.

%
%



\begin{figure}
\epsscale{1.15}
\plotone{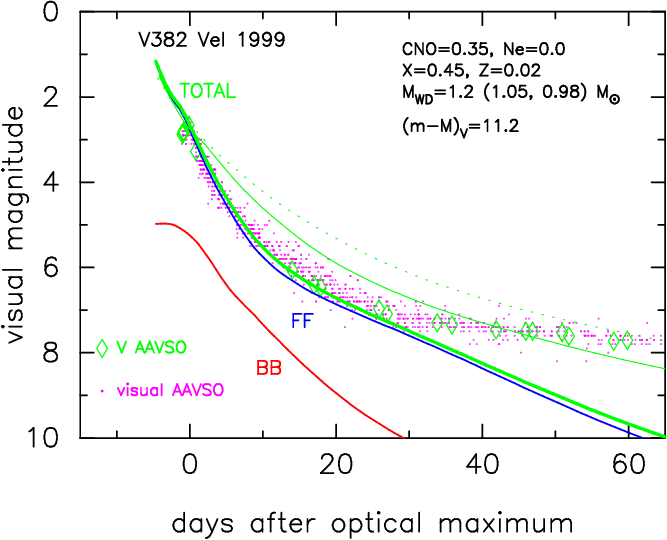}
\caption{
Optical visual (magenta dots) and $V$ (open green diamonds)
magnitudes of the classical nova V382 Vel are plotted on a linear timescale.
The visual and $V$ data are taken from AAVSO.
We assume the $V$-band distance modulus of $(m-M)_V= 11.2$ to the nova,
i.e., $A_V + 5 \log (d / 10 {\rm ~pc}) 
= 0.3 + 5 \log (1.5 {\rm ~kpc} / 10 {\rm ~pc}) = 11.2$.
We plot our best-fit model of 1.2 $M_\sun$ WD, that is,
the red line denotes the blackbody emission from the photosphere
(labeled BB),
the blue line is the free-free emission light curve (labeled FF), and
the green line corresponds to the total flux of FF and BB (labeled TOTAL).
Only the decay phase of the steady-state envelope model is plotted.
We also plot the total emission (BB+FF) model light curves for
another two WD mass models, 1.05 $M_\sun$ (thin solid green line)
and 0.98 $M_\sun$ (thin dotted green line).  We select the 1.2 $M_\sun$
model as the best-fit model among the three mass models with the chemical
composition of CO3.
\label{optical_mass_v382_vel_x45z02c15o20}}
\end{figure}


\begin{figure}
\gridline{\fig{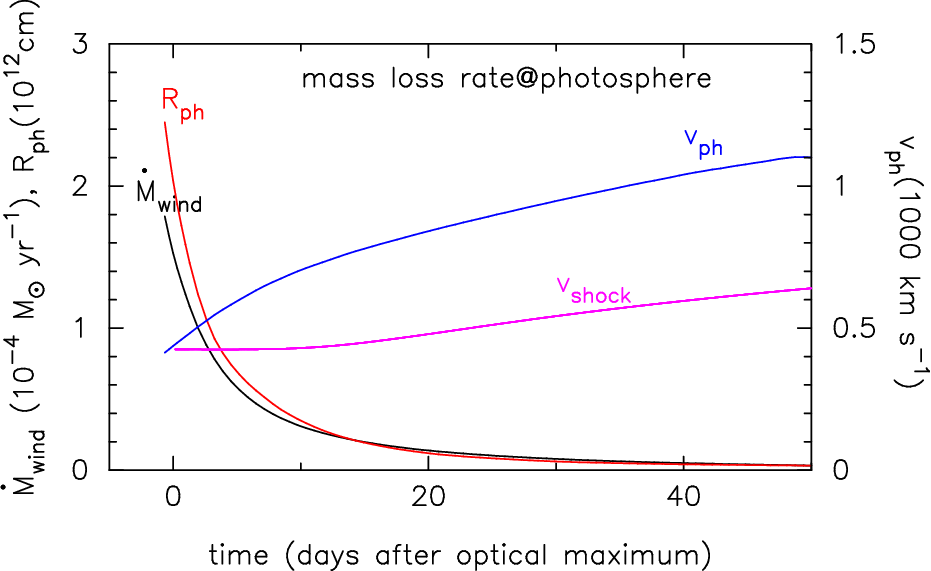}{0.47\textwidth}{(a)}
          }
\gridline{
          \fig{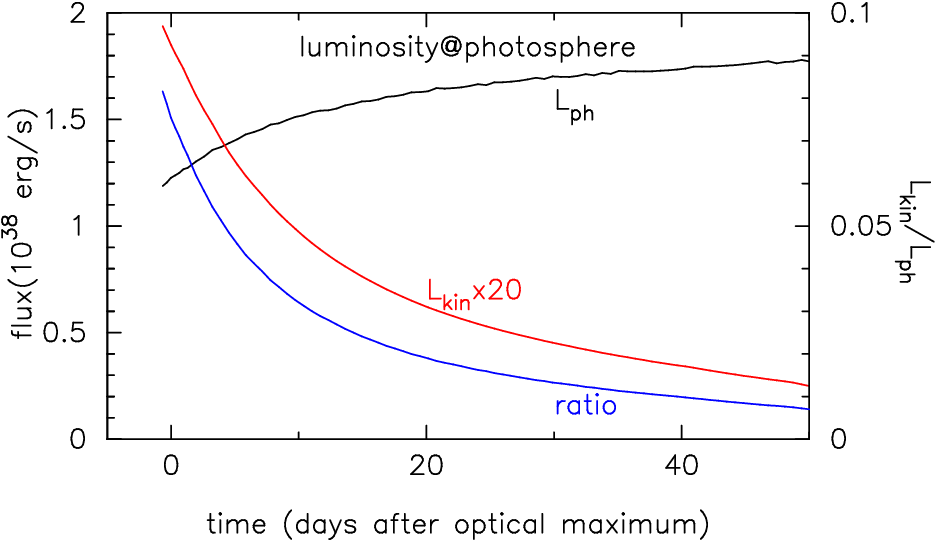}{0.47\textwidth}{(b)}
          }
\caption{
Same as those in Figure \ref{mdot_radius_velocity}, but for 
a $1.2~M_\sun$ WD with CNO enhancement (CO3).
Only the decay phase (after the optical maximum) of steady-state envelope
model is plotted.
The time $t=0$ corresponds to the optical maximum. 
\label{mdot_radius_velocity_v5856_sgr}}
\end{figure}


\begin{figure}
\gridline{\fig{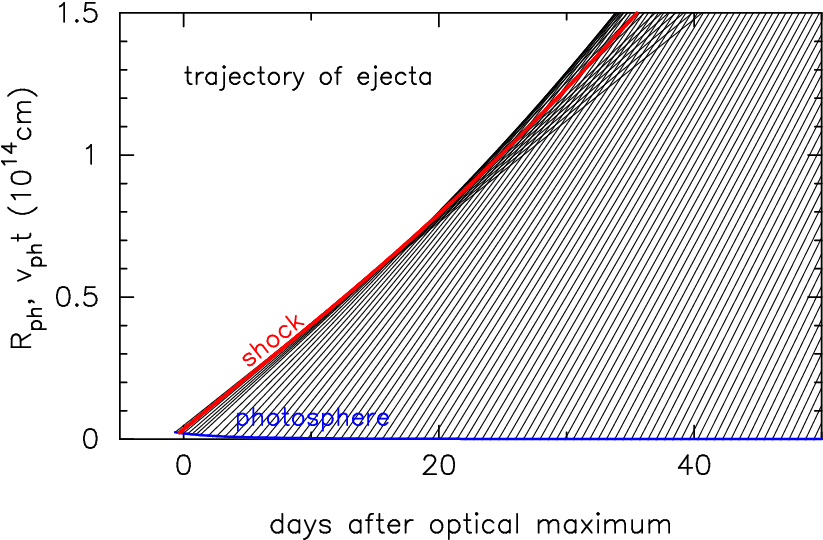}{0.47\textwidth}{(a)}
          }
\gridline{
          \fig{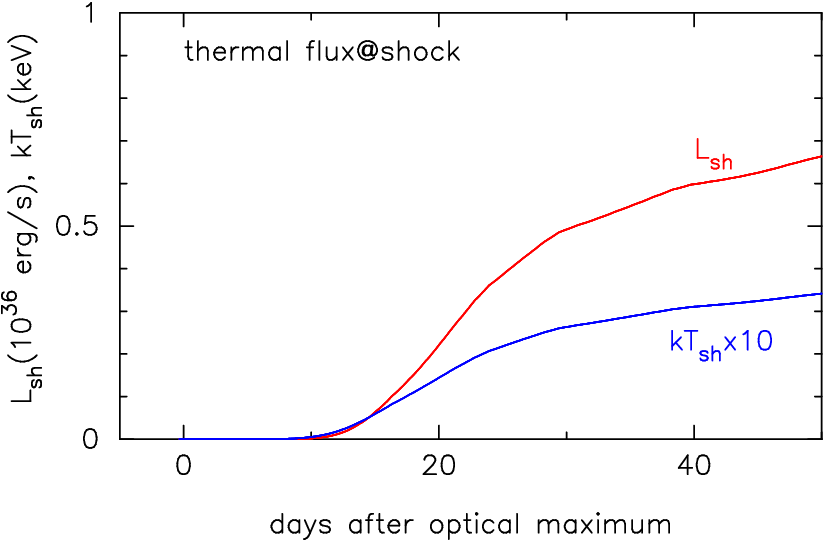}{0.47\textwidth}{(b)}
          }
\gridline{
          \fig{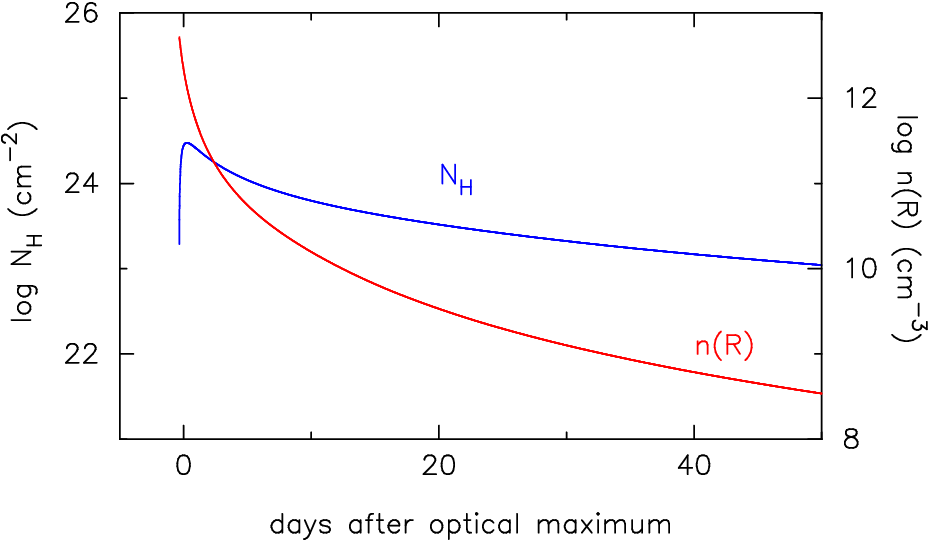}{0.47\textwidth}{(c)}
          }
\caption{Same as in Figure \ref{shock_density_pp_interaction},
but for a $1.2~M_\sun$ WD with CNO enhancement (CO3).
Only the decay phase (after the optical maximum) of the steady-state
envelope model is plotted.
{\bf (a)} A strong shock arises soon after the optical maximum
and expands outward at a speed of 400--700 km s$^{-1}$.
{\bf (b)} Both the shocked luminosity $L_{\rm sh}$ (red)
and temperature $k T_{\rm sh}$ (blue)
quickly rise up $\sim 15$ days after the optical maximum.
{\bf (c)} The $N_{\rm H}$ is still as high as a few times 
$10^{23}$ cm$^{-2}$ at $t\sim 20$ days after the optical maximum.
\label{density_pp_interaction_v5856_sgr}}
\end{figure}

\subsection{V382 Vel 1999}
\label{v382_vel_1999}

\subsubsection{Model Light Curve Fitting of V382 Vel}
\label{model_fitting_v382_vel_wd_mass}

Figure \ref{optical_mass_v382_vel_x45z02c15o20} shows our 
model light curve fitting \citep[see, e.g.,][for more details]{hac19kb}.
We adopt a 1.2 $M_\sun$ WD model
as the best-fit one among the three, 0.98, 1.05, and 1.2 $M_\sun$.
Here, we assume a $V$-band distance modulus of
$(m-M)_V= A_V + 5 \log (d/10{\rm ~pc})= 0.3 + 
5 \log (1.5{\rm ~kpc}/10{\rm ~pc}) = 11.2$, where a relatively low
extinction $A_V\approx 0.3$ is taken from \citet{del02pd}
and the distance is taken from Gaia EDR3 \citep{bai21rf}.

The $V$-band distance modulus is fixed to be $(m-M)_V= 11.2$, so that
we have only one degree of freedom, that is, shifting the model light curves
back and forth horizontally. In the figure, we fit our three (total $V$)
model light curves with the optical maximum.  The model light curve
labeled TOTAL reasonably follows the visual and $V$ data of V382 Vel
in an early phase (up to $\sim 30$ days after the optical maximum).
After that, the model light curve deviates downward 
largely from the visual and
$V$ data.  This is because strong emission lines such as [\ion{O}{3}]
significantly contribute to the $V$-band flux as already discussed
in section \ref{model_fitting_v1974_cyg_wd_mass}.
Thus, the continuum flux of the nova decreases almost along with
the model light curve.
Therefore, we adopt the 1.2 $M_\sun$ as the best-fit one among the three
model light curves.

The optically thick wind of this 1.2 $M_\sun$ WD model ends at 
$t\sim 95$ days after the optical maximum.  The hydrogen shell burning
stops at $t\sim 190$ days after the optical maximum. These features
are broadly consistent with the decay of supersoft X-ray
phase observed with BeppoSAX \citep{ori02pg}. See 
\citet{hac16k} for more details.

\subsubsection{Shock Formation outside the Photosphere}
\label{shock_outside_photosphere_v382_vel}

Figures \ref{mdot_radius_velocity_v5856_sgr} and
\ref{density_pp_interaction_v5856_sgr} show the same physical
quantities as in Figures \ref{mdot_radius_velocity} and 
\ref{shock_density_pp_interaction}, respectively,
but for a 1.2 $M_\sun$ (CO3) model.
A strong shock arises outside the photosphere and expands
at a velocity of $v_{\rm sh}(=v_{\rm shock})
\approx 400$ -- 700 km s$^{-1}$
as shown in Figures \ref{mdot_radius_velocity_v5856_sgr}a and
\ref{density_pp_interaction_v5856_sgr}a.
The main differences from the $0.98 ~M_\sun$ model are
(i) the wind velocity is faster in the 1.2 $M_\sun$,
400 (versus 250) km s$^{-1}$ soon after the optical maximum, and 
1100 (versus 1000) km s$^{-1}$ in a later phase; as a result,
(ii) the shock velocity is faster in the 1.2 $M_\sun$,
400 (versus 250) km s$^{-1}$ soon after optical maximum, but similar to
700 (versus 600) km s$^{-1}$ in a later phase;
(iii) the duration of wind phase is much shorter
in the 1.2 $M_\sun$, 95 days (versus 250 days).
Thus, the evolution timescale is about 2.5 times shorter 
in the 1.2 $M_\sun$ model.


\subsubsection{Velocity Systems in V382 Vel}
\label{velocity_systems_v382_vel}

The temperature just behind the shock is obtained to be
$\sim 0.03$ keV as shown in Figure
\ref{density_pp_interaction_v5856_sgr}b.
This temperature is much smaller than those estimated 
from hard X-ray emission by \citet{muk01i}.
The velocities of the principal and diffuse-enhanced systems
in V382 Vel were obtained by \citet{del02pd} to be
$v_{\rm p}\sim$ 2300 km s$^{-1}$ and $v_{\rm d}\sim$ 3700 km s$^{-1}$
at post-maximum.  Our $v_{\rm shock} = v_{\rm p}\sim 400$ km s$^{-1}$ 
is about 6 times smaller than that of V382 Vel. 
Simply substituting the observed values $v_{\rm wind} - v_{\rm shock}
= v_{\rm d} - v_{\rm p} = 3700 - 2300 = 1400$ km s$^{-1}$,
we obtain $k T_{\rm sh}\sim 2$ keV,
which is consistent with the temperature
estimated by \citet{muk01i}.

\subsubsection{Hard X-Ray Emission from V382 Vel}
\label{hard_x-ray_emission_v382_vel}

Hard X-ray emission from V382 Vel was detected with ASCA and
RXTE from about 20 days to 60 days past the optical peak \citep{muk01i}. 
%
%
%
The column density of hydrogen is estimated from equation
(\ref{column_density_hydrogen_time}) together with
$v_{\rm shock} = v_{\rm p} = 2300$ km s$^{-1}$.
We obtain the ejecta mass of $M_{\rm ej}\sim 0.51 \times 10^{-5} ~M_\sun$
from the light curve fitting in Figure
\ref{optical_mass_v382_vel_x45z02c15o20}.
Substituting these values into equation (\ref{column_density_hydrogen_time}),
we have $N_{\rm H}= 1.3\times 10^{21}$ cm$^{-1}$ at $t\sim 20$ day
and $N_{\rm H}= 2\times 10^{20}$ cm$^{-1}$ at $t\sim 50$ day, both
of which are low enough for X-ray photons to diffuse out from the shell.

The thermal energy generation rate is as large as
$L_{\rm sh}\sim 1.8 \times 10^{37} \times (1.4)^3/3.7 =2.6 \times 10^{37}$
erg s$^{-1}$ from equation (\ref{shocked_energy_generation})
for $\dot{M}_{\rm wind}=2\times 10^{-4} ~M_\sun$ yr$^{-1}$ near post-maximum,
which decreases to $L_{\rm sh}\sim 2.7 \times 10^{36}$ erg s$^{-1}$ 
for $\dot{M}_{\rm wind}=2\times 10^{-5} ~M_\sun$ yr$^{-1}$ 
at $\sim 50$ day after the maximum. 
Here, with a relation of $t_{\rm ret}=t\times (v_{\rm p}/v_{\rm d})$,
we have $t_{\rm ret}\sim t\times (2300/3700)=0.62~t$, that is, 
$t_{\rm ret}= 50\times 0.62 =31$ day.  If we take the retarded (look back)
time $t_{\rm ret}$ into account, 
the wind mass-loss rate measured at the photosphere
decreases to $\dot{M}_{\rm wind}= 2.0\times 10^{-5}
~M_\sun$ yr$^{-1}$ in our 1.2 $M_\sun$ WD (CO3) model at $+19$ day.
The shock luminosity $L_{\rm sh}\sim 2.7 \times 10^{36}$ erg s$^{-1}$
is sufficiently larger than $L_X \sim 7 \times 10^{34}$ erg s$^{-1}$
estimated by \citet{muk01i} at $\sim 50$ day after maximum.
Then the conversion rate to X-ray emission is rather small, i.e.,
$L_X/L_{\rm sh}\sim 0.026$.

Thus, our shock model of a nova reasonably explains
the hard X-ray emission from V382 Vel.

\begin{figure*}
\gridline{\fig{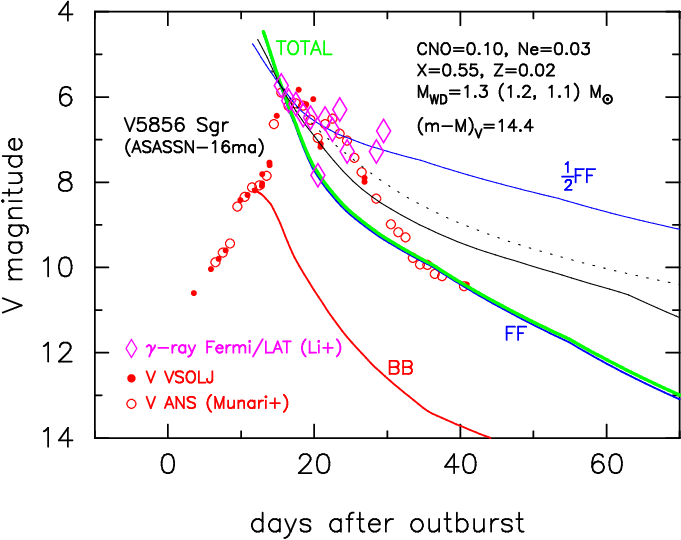}{0.47\textwidth}{(a)}
          \fig{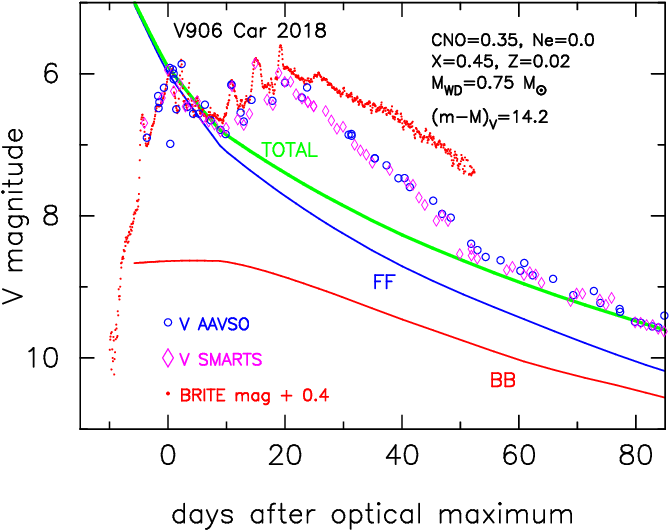}{0.47\textwidth}{(b)}
          }
\gridline{\fig{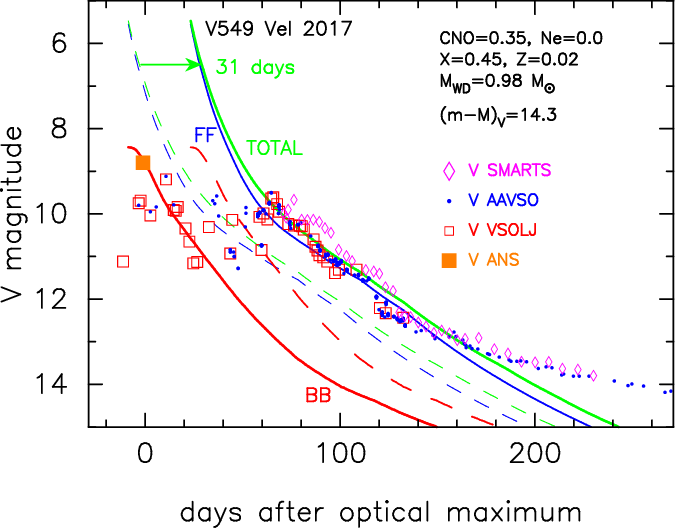}{0.47\textwidth}{(c)}
          \fig{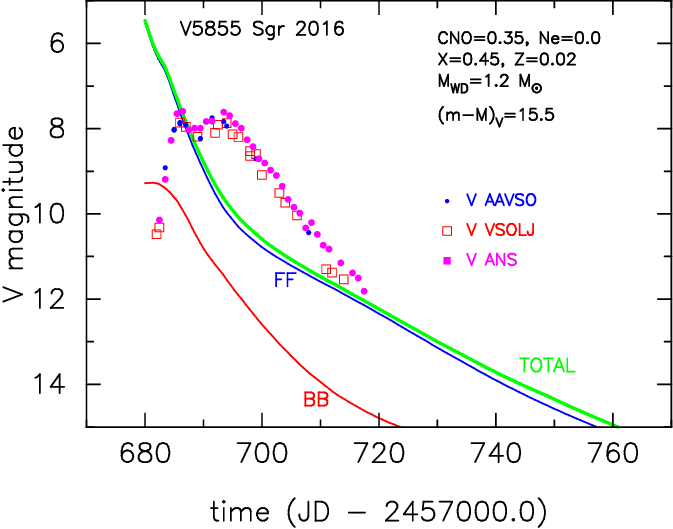}{0.47\textwidth}{(d)}
          }
\caption{{\bf (a)} 
Optical $V$ magnitudes of the classical nova V5856 Sgr
(ASASSN-16ma) are plotted on a linear timescale.
The $V$ data are taken from \citet{mun17hf} and the 
Variable Star Observers League of Japan (VSOLJ).
We assume the $V$-band distance modulus of $(m-M)_V= 14.4$ to the nova.
The red line denotes the blackbody emission from the photosphere
(labeled BB) of a 1.3 $M_\sun$ WD (Ne2) model.
The blue line is the free-free emission light curve (labeled FF)
of the 1.3 $M_\sun$ WD.
The green line corresponds to the total flux of FF and BB (labeled TOTAL)
of the 1.3 $M_\sun$ WD.
We added two other WD mass models of total $V$,
1.2 $M_\sun$ (thin solid black line)
and 1.1 $M_\sun$ (thin dotted black line).
We also added the gamma-ray fluxes observed with the Fermi/LAT
\citep[open magenta diamonds;][]{li17mc} and half a slope of free-free
emission light curve (thin blue line labeled ${1 \over 2}$FF).  
{\bf (b)} 
The $V$ light curve of V906 Car. 
The optical $V$ data are taken from AAVSO and SMARTS \citep{wal12},
and the BRITE data are from \citet{ayd20sc}.
We assume $(m-M)_V= 14.2$.
We fit a $0.75 ~M_\sun$ WD (CO3) model to the $V$ light curve.
{\bf (c)}
The $V$ light curve of V549 Vel.
The optical $V$ data are taken from SMARTS \citep{wal12},
AAVSO, and VSOLJ. The first peak ($V=8.8$) is taken from 
ANS collaboration \citep[filled orange square;][]{li20hm}, and 
we set $t=0$ day at this first $V$ peak (MJD 58032.0).
We assume $(m-M)_V= 14.3$.
We fit a $0.98 ~M_\sun$ WD (CO3) model to the $V$ light curve.
The first and second jitters are fitted with the red line labeled BB,
while the last jitter and its decay phase are
fitted with the green line labeled TOTAL, which is separated by 31 days
from the first model light curve.
See the text for more details.
{\bf (d)}
The $V$ light curve of V5855 Sgr.
The optical $V$ data are taken from
AAVSO, VSOLJ, and ANS \citep{mun17hf}.
We assume $(m-M)_V= 15.5$.
We adopt a $1.2 ~M_\sun$ WD (CO3) model to fit it with the $V$ light curve.
\label{optical_v5856_sgr_v906_car_v549_vel_v5855_sgr}}
\end{figure*}

\subsection{ASASSN-16ma (V5856 Sgr 2016)}
\label{comparison_asassn16ma_v5856sgr}

\citet{li17mc} reported Fermi/LAT observation of 
the classical nova ASASSN-16ma (a.k.a. V5856 Sgr).
Its dense observation provides high-quality information.  
We summarize their results and implications as follows.
\begin{enumerate}
\item  The first gamma-ray emission is detected shortly 
after the optical maximum. They have no positive detection 
before the optical maximum. 
\item The gamma-ray flux decays to almost follow the decline trend of
the optical flux. The ratio of the gamma-ray flux to the optical flux
is almost constant, $L_\gamma / L_{\rm opt} \sim 0.002$. 
\item From a remarkable correlation between the $\gamma$-ray and 
optical fluxes, they concluded that 
the majority of the optical light comes from 
reprocessed emission from shocks rather than the WD.
\item Their assumed model is such that:
after the thermonuclear runaway, the WD loses its mass by slow winds 
that escape from the binary through the outer Lagrangian (L2) point
to form circumbinary matter.  Then, the wind becomes faster and
collides with the circumbinary matter. 
\item A strong shock arises deep inside the circumbinary 
matter, where gas is optically thick for X-rays so that all the 
X-ray photons are reprocessed to lower-wavelength photons
(UV and optical).  GeV gamma-rays diffuse out from the shocked layer
because its optical depth is small.
\end{enumerate}
A similar property to their point 1 was reported in 
two classical novae with a short orbital period,
V339 Del \citep[$P_{\rm orb}= 3.154$ hr in][]{cho15sk}
and V1324 Sco \citep[$P_{\rm orb}= 3.8$ hr in][]{fin18cm};
the Fermi/LAT $> 100$ MeV flux peaked a few to several days
after the optical maximum \citep[see Figure 3 of][]{che15jsg}.

For their point 2, no clear correlation is seen
in the V549 Vel outburst \citep{li20hm}
and YZ Ret outburst \citep{kon22wa, sok22ll}. 
So it may not be common among all the gamma-ray novae.

The time sequence of the GeV gamma-ray and optical light curves
are as follows:
\begin{itemize}
\item The nova optically brightens up to $m_V \sim 8$ over two weeks, and 
there is no detection of GeV gamma-rays during this rising phase (Phase I).
\item Then its brightness rapidly increased by a factor of 10 in two days, 
reaching an optical maximum of $m_V \sim 5.4$ (Phase II).  GeV
gamma-rays were detected soon after this optical maximum.
\item The gamma-ray flux declined with being well correlated to
the decay of the optical flux until 9 days after the optical maximum
(Phase III).  They concluded that the duration of detectable
gamma-ray emission is $9 - 15$ days.
\item No hard X-rays (1-10 keV) were detected by Swift/XRT
around the optical maximum or five months later.
\end{itemize}
The optical and gamma-ray light curves are plotted 
in Figure \ref{optical_v5856_sgr_v906_car_v549_vel_v5855_sgr}a.
We discuss GeV gamma-ray fluxes from V5856 Sgr,
assuming that $v_{\rm shock} = v_{\rm p}$ and
$v_{\rm wind}= v_{\rm d}$.

\subsubsection{Model Light Curve Fitting}
\label{model_fitting_wd_mass}

Figure \ref{optical_v5856_sgr_v906_car_v549_vel_v5855_sgr}a shows our 
model light curve fitting.  We adopt a 1.3 $M_\sun$ WD model
as the best-fit one among the three, 1.3, 1.2, and 1.1 $M_\sun$, but
for the envelope chemical composition of neon nova 2 (Ne2), i.e.,
$X=0.55$, $Y=0.30$, $Z=0.02$, $X_{\rm O}=0.10$, and $X_{\rm Ne}=0.03$ 
\citep[e.g.,][]{kat94h, hac06kb, hac10k}.
This is because such a heavy WD is composed of an oxygen/neon core
\citep[e.g.,][]{ume99ny}.
Here, we assume a $V$-band distance modulus of
$(m-M)_V= A_V + 5 \log (d/10{\rm ~pc})= 14.4$.  This value is
an arithmetic mean between those of \citet{li17mc} and \citet{mun17hf}.

Because the $V$-band distance modulus is fixed to be $(m-M)_V= 14.4$,
we have only one freedom, that is, shifting the model light curves back and
forth horizontally. In the figure, we fit our three (total $V$)
model light curves with the optical maximum.  Although the midway
data points are slightly above the model line, the late data
around 40 days after the optical maximum are just on the model line (thick
green line).  Therefore, we adopt the 1.3 $M_\sun$ as the best-fit one.

We also plot the gamma-ray fluxes (open magenta diamonds) in 
the figure. Qualitatively, the trend of gamma-ray decay follows
well the trend of the optical $V$ decay.  In our model, the flux of free-free
emission dominates and is given by $L_{V, \rm ff}\propto 
(\dot{M}_{\rm wind}/v_{\rm ph})^2$ while the flux of the gamma-rays is given
by $L_{\gamma, \rm sh} \propto (\dot{M}_{\rm wind}/v_{\rm wind})$.
Thus, the dependency on $(\dot{M}/v)^n$ is similar, 
but the power of $n$ is different by a factor of 2 ($n=2$ versus $n=1$).
Therefore, the light curves in magnitudes are written as
\begin{eqnarray}
{1 \over 2} m_{V,\rm ff} &=& 
{1\over 2}(-2.5 \log L_{V, \rm ff}) + {\rm ~const.} \cr
&=& -2.5 \log (\dot{M}/v) + {\rm ~const.} \cr
&=& -2.5 \log L_{\gamma, \rm sh} + {\rm ~const.} \cr
&=& m_{\gamma, \rm sh} + {\rm ~const.},
\end{eqnarray}
so the slope of decay is different by a factor of 2.
We plot the $m_{\gamma, \rm sh}$ (thin blue line labeled ${1 \over 2}$FF) in 
Figure \ref{optical_v5856_sgr_v906_car_v549_vel_v5855_sgr}a, which
represents the decay trend of GeV gamma-ray fluxes rather than the optical
$V$ flux itself.

\subsubsection{Eddington Luminosity in the Fireball Stage}
\label{fireball_stage}

\citet{mun17hf} concluded that Phase I of \citet{li17mc} ($t < 15$ days
in Figure \ref{optical_v5856_sgr_v906_car_v549_vel_v5855_sgr}a)
is the fireball stage of a nova.  This means that 
we did directly observe the photosphere of the nova, 
so the observed magnitude is a direct result of the theoretical 
$L_{\rm ph}$, and not of the summation of $L_{\rm ph}$ and  
$L_{\rm ff}$.
This occurs if free-free emission becomes dominant 
only after the optical peak, as suggested by \citet{hac06kb}. 
The sharp increase from $V \sim 8.4$ to $V \sim 6$
just before the optical maximum can be explained
as a sudden increase in the free-free emission. 

Our theoretical photospheric luminosity $L_{\rm ph}$, which is
close to the Eddington limit, 
corresponds to the absolute $V$ magnitude of $M_V\approx -6.2$. 
This value is converted to the apparent $V$ magnitude of 
$m_V = (m-M)_V + M_V = 14.4 - 6.2 = 8.2$
at $t \sim 15$ day (the last days of Phase I). 
Figure \ref{optical_v5856_sgr_v906_car_v549_vel_v5855_sgr}a 
supports our interpretation that the $V$ brightness jumps 
from the photospheric luminosity close to the Eddington limit
to a phase dominated by free-free emission.

This is a possible theoretical explanation for the pre-maximum halt,
which was already pointed out by \citet{hac04k}.
They proposed a relation between the WD mass and $V$ brightness
at the pre-maximum halt, i.e., $M_{V, \rm halt}\approx
-1.75 (M_{\rm WD}/M_\sun) - 4.25$, for the chemical
composition\footnote{The photospheric Eddington luminosity,
$L_{\rm Edd, ph}= 4 \pi cGM_{\rm WD}/\kappa_{\rm ph}$,
depends on the opacity at the photosphere, $\kappa_{\rm ph}$.
The opacity is not only
a function of $T$ and $\rho$ but also depends on the chemical composition.} 
of C$+$O=0.30 (CO2), which gives $M_V\sim -6.5$ for 1.3 $M_\sun$ WD (CO2),
slightly brighter than our $M_V\sim -6.2$ for the chemical composition
of Ne2.  

To summarize, we directly observe the photosphere
so that we get only the photospheric luminosity $L_{\rm ph}$ at
the pre-maximum halt.  
After that, free-free emission dominates, and the luminosity can 
exceed the Eddington limit.

\subsubsection{Shock outside the Photosphere}
\label{shock_outside_photosphere}

\citet{li17mc} reported the P Cygni absorption profiles on H$\alpha$ line
in their Supplementary Figure 2.
We adopted $v_{\rm p}= 800$ km s$^{-1}$ and $v_{\rm d}= 2200$ km s$^{-1}$
as already obtained in section \ref{velocity_system_v5856_sgr}.
Then, the shocked thermal temperature
is k$T_{\rm sh}= 2.0$ keV from equation (\ref{shock_kev_energy}),
which is high enough to emit hard ($\gtrsim 1$ keV) X-rays.
The shocked thermal energy generation rate is estimated to be
$L_{\rm sh}= 1.8\times 10^{37} (1.4)^3/2.2 \times 1.0 = 2.2\times 10^{37}$
erg s$^{-1}$ from equation (\ref{shocked_energy_generation})
for $\dot {M}_{\rm wind}= 1.0\times 10^{-4} ~M_\sun$ yr$^{-1}$
from our model light curve fitting.

GeV gamma rays were detected from $+1$ day to $+14$ day (after the optical
maximum).  The optical depth of gamma rays ($\sim$ GeV) is
\begin{eqnarray}
\tau_\gamma & \approx & 8\times 10^{-3} \left( {{{\dot M}_{\rm wind}} \over
{10^{-4} M_\sun ~ {\rm yr}^{-1}}} \right)
\left( {{r} \over {10^{13} ~{\rm cm}}} \right)^{-1} \cr
& & \times 
\left( {{v_{\rm wind}} \over {400 ~{\rm km} ~ {\rm s}^{-1}}} \right)^{-1}
\left( {{E_X} \over {{\rm 300~MeV}}} \right)^{-1},
\label{gev_gammaray_optical_depth}
\end{eqnarray}
which is taken from Equation (11) of \citet{li17mc}.
This is small enough for gamma rays to diffuse out from the shocked
matter, even if we adopt 
${\dot M}_{\rm wind}\sim 10^{-3} ~M_\sun$ yr$^{-1}$
and $v_{\rm wind}\sim 4000$ km s$^{-1}$ \citep[see, e.g.,][]{mar18dj}.

\citet{li17mc} estimated the GeV gamma-ray luminosity to be
$L_\gamma \sim 1\times 10^{36}$ erg s$^{-1}$, which gives
the conversion rate to GeV gamma rays, $L_\gamma / L_{\rm sh} \sim 0.04$,
in our shock model.

\subsubsection{Hard X-Ray Emission}
\label{hard_x-ray_emission_v5856_sgr}

The ejecta mass is calculated to be $M_{\rm ej}=0.25 \times 10^{-5}$
$M_\sun$ from our 1.3 $M_\sun$ WD model. Then, we obtain the emergence
time of hard X-rays of $t_{\rm H21} \sim 38$ days for $v_{\rm p}= 1000$
km s$^{-1}$.  Here, the velocity of the principal system increased to
$v_{\rm p}= 1100$ km s$^{-1}$ $11$ days after the optical maximum
\citep[see Supplementary Figure 2 of][]{li17mc}, so we adopt 
$v_{\rm p}=1000$ km s$^{-1}$ for an averaged value from 
$-500$ km s$^{-1}$ to $-1100$ km s$^{-1}$ for the principal system.
If we set $N_{\rm H}\sim 10^{22}$ cm$^{-2}$ for hard X-ray detection
\citep{muk01i}, we have $t_{\rm H22}\sim 10$ days
for $M_{\rm ej}=M_{\rm shell}\sim 0.15 \times 10^{-5} ~M_\sun$. 
However, no hard X-ray emission from ASASSN-16ma (V5856 Sgr)
was detected with Swift/XRT both
around the optical peak and five months later \citep{li17mc}. 
This is because the optical depth for hard X-rays is too large
($\tau_X \sim 10^3$) post-maximum, and optically thick 
winds had already stopped at $t \sim 74$ day after the optical maximum
in our 1.3 $M_\sun$ WD (Ne2) model.
In other words, the shock wave has already faded out before
$t \sim 135$ day ($\sim $five months) 
because the wind had stopped much earlier than this epoch.
We will revisit the reason of no detection of hard and soft X-rays
in section \ref{no_hard_x-ray_detection_v5856_sgr} in more detail.

%

\subsubsection{Optical Depth to a Shocked Layer}
\label{opticaly_depth_shock_v5856_sgr}

Optical depth to a shocked layer is estimated to be
\begin{eqnarray}
\tau_{\rm opt} & \approx & 0.42 \left( {{{\dot M}_{\rm wind}} \over
{10^{-4} M_\sun ~ {\rm yr}^{-1}}} \right)
\left( {{r} \over {10^{13} ~{\rm cm}}} \right)^{-1} \cr
& & \times 
\left( {{v_{\rm wind}} \over {400 ~{\rm km} ~ {\rm s}^{-1}}} \right)^{-1}
\left( {{\kappa_{\rm opt}} \over {0.1 {\rm ~cm}^{-2} {\rm ~g}^{-1}}} \right),
\label{optical_depth_shocked_layer}
\end{eqnarray}
which is taken from equation (7) of \citet{li17mc}.
Here, $\kappa_{\rm opt}$ is the opacity.
If we adopt $v_{\rm wind}= 1000$ (or 400) km s$^{-1}$ and
a shock radius of $r\sim 10^{14}$ cm, we have $\tau_{\rm opt}
\sim 10^{-2}$ (or $\sim 10^{-1}$).
Thus, the shocked layer is always optically thin.  In other words,
we can always see the photosphere and inner wind through the shocked
layer as illustrated in Figure \ref{wind_shock_config}.

\subsection{Other Novae Detected in G{\lowercase{e}}V Gamma-Rays}
\label{other_gev_gamma-ray_novae}

What makes a nova observable in GeV gamma-rays? 
Equation (\ref{shocked_energy_generation}) gives a hint. 
The shocked energy flux depends on the velocity difference 
between $v_{\rm p}$ and $v_{\rm d}$ at post-maximum as well 
as the wind mass-loss rate $\dot{M}_{\rm wind}$, which 
is the largest at the optical maximum and decreases afterward.
In our steady-state optically thick wind model, the wind mass-loss
rate near the optical peak is as large as $(1-3) \times 10^{-4} ~M_\sun$
yr$^{-1}$ (see Figures \ref{mdot_radius_velocity},
\ref{mdot_radius_velocity_v1668_cyg}, and
\ref{mdot_radius_velocity_v5856_sgr}) and decreases afterwards.   
We regard that the decay in the gamma-ray flux is governed by the decrease
in the wind mass-loss rate, as shown by the ${1 \over 2}$FF line in Figure 
\ref{optical_v5856_sgr_v906_car_v549_vel_v5855_sgr}a, rather than
by the decrease in the velocity difference $v_{\rm d}-v_{\rm p}$. 

In this subsection, 
we analyze three novae, V906 Car, V549 Vel, and
V5855 Sgr, whose ejecta velocities are well observed
at post-maximum.  All the spectroscopic data are taken from \citet{ayd20ci}.

\subsubsection{V906 Car 2018}
\label{v906_car_2018}

V906 Car is a nova detected in GeV gamma-rays \citep{ayd20sc}.
This nova shows a remarkable correlation between the optical and
gamma-ray fluxes, like in V5856 Sgr \citep{li17mc, ayd20sc}.

\citet{ayd20sc} estimated the distance and reddening to V906 Car
to be $d\sim 4.0$ kpc and $A_V= 1.2$ mag (or $E(B-V)=A_V/3.1= 0.4$ mag),
respectively, so we have the distance modulus in the $V$ band of
$(m-M)_V= A_V + 5 \log (d/{\rm 10 ~pc}) = 14.2$.
Figure \ref{optical_v5856_sgr_v906_car_v549_vel_v5855_sgr}b shows
the $V$ light curve of V906 Car together with BRITE magnitudes,
the data of which are taken from AAVSO, SMARTS, and \citet{ayd20sc}.
The BRITE filter transmission is between 550 nm and 700 nm.

The optical spectra in an early phase were obtained by \citet{ayd20ci}.
They presented the velocities of the deepest narrow P Cygni absorption lines
on H$\alpha$, which increase from $-210$ km s$^{-1}$ at the optical
maximum (day 0) to $-340$ km s$^{-1}$ at 6 days after the maximum (day 6).
We adopt these $-210$ km s$^{-1}$ and $-340$ km s$^{-1}$ as the velocities
of the principal system postmaximum (at $+0$ day and $+6$ day, respectively).
A broad absorption feature from $-1800$ km s$^{-1}$ to $-1200$ km s$^{-1}$ 
was observed at day 6 in the same H$\alpha$ line.
We assign this broad velocity feature to the velocity of diffuse-enhanced
system postmaximum (at $+6$ day).
Substituting $v_{\rm p}= (210+340)/2=275$ km s$^{-1}$ 
and $v_{\rm d}= (1800+1200)/2= 1500$ km s$^{-1}$ into 
equation (\ref{shocked_energy_generation}), we obtain 
$L_{\rm sh}\sim 1.8 \times 10^{37} \times (1.23)^3/1.5 =2.2 \times 10^{37}$
erg s$^{-1}$ for $\dot{M}_{\rm wind}=1\times 10^{-4} ~M_\sun$ yr$^{-1}$. 

\citet{ayd20sc} estimated the GeV gamma-ray flux of V906 Car to be
$L_\gamma \sim 1\times 10^{36}$ erg s$^{-1}$ from day 14 to day 34,
which amounts to 5\% of our shock luminosity $L_{\rm sh}$, i.e.,
$L_\gamma / L_{\rm sh}= 0.05$.  There are no Fermi/LAT
observations between day 0 and day 14 because Fermi/LAT was down.

It should be noted that we must increase the wind mass-loss rate
from $\dot{M}_{\rm wind}\sim 1 \times 10^{-4} ~M_\sun$ yr$^{-1}$
to $\dot{M}_{\rm wind}\sim 2\times 10^{-3} ~M_\sun$ yr$^{-1}$ or more
if we take a ratio of $L_\gamma / L_{\rm opt}\sim 0.001$--0.002 and
the shock luminosity covers all the optical luminosity, which
amounts to $L_{\rm opt}\sim 10^{39}$ erg s$^{-1}$, as suggested
by \citet{li17mc} and \citet{ayd20sc}. 
These estimates are based on the very good correlation between
the gamma-ray and optical fluxes.  Such a large mass-loss rate,
however, challenges our optically thick wind model \citep{kat94h}. 



\citet{sok20mc} reported the NuSTAR observation
$+24$ and $+45$ days after the optical maximum.
They obtained the temperature, k$T\sim 8.6$ keV and 4.4 keV,
respectively, at day 24 and day 45, the hydrogen column density,
$N_{\rm H}\sim 4.3\times 10^{22}$ cm$^{-2}$ and $0.6\times 10^{22}$
cm$^{-2}$.
The Swift/XRT detected X-rays many times from day $\sim 50$ to day 440.
The temperature of optically thin plasma decreases from several keV to
a few tenths keV as well as the hydrogen column density
from $0.6 \times 10^{22}$ cm$^{-2}$ to $0.01 \times 10^{22}$ cm$^{-2}$
\citep{sok20mc}.

We have estimated the hydrogen column density $N_{\rm H}$ at the two epochs.
The shocked shell mass is $M_{\rm shell}= 0.85 \times 10^{-5} ~M_\sun$
(at day 24) and $M_{\rm shell}= 1.27 \times 10^{-5} ~M_\sun$ 
(at day 45) calculated from our $0.75 ~M_\sun$ model (it reached
$M_{\rm shell}= 2.5\times 10^{-5} ~M_\sun$ when the winds stopped).
Substituting these values together with the shell velocity (= shock
velocity) of $v_{\rm p}= 420$ km s$^{-1}$ \citep[at day 
16.5 in Figure 11 of ][]{ayd20ci}
and $v_{\rm p}\sim 760$ km s$^{-1}$ (at day 40,
linear extrapolation from their Figure 11) 
into Equation (\ref{column_density_hydrogen_time}),
we obtain $N_{\rm H}\sim 4.8\times 10^{22}$ cm$^{-2}$ at day 24
and $0.63\times 10^{22}$ cm$^{-2}$ at day 45, being broadly consistent
with \citet{sok20mc}'s values.

We have estimated k$T_{\rm sh}\approx ((1800 - 340)/1000)^2 \sim 2$ keV
at day 6 from equation (\ref{shock_kev_energy}) by using higher-velocity
component of the diffuse-enhanced system.  This value is much lower than
their estimates based on the NuSTAR data, but consistent with
the temperature based on the Swift data between day 168 and day 238
\citep{sok20mc}. 

Our 0.75 $M_\sun$ WD model predicts the wind duration of 
$t_{\rm wind}\sim 720$ days and the hydrogen shell burning
up to $\sim 2360$ days, although these durations
depend on the chemical composition of the envelope, $X$ and
$X_{\rm CNO}$.  If this is the case, supersoft X-ray emission
emerges about two years after the outburst and 
continues up to 6.5 years after the outburst at a rate of
$L_{\rm ph}\sim 7\times 10^{37}$ erg s$^{-1}$ and
k$T_{\rm ph}\sim 40$ eV \citep{kat94h, hac06kb, hac10k}.

%

\subsubsection{V549 Vel 2017}
\label{v549_vel_2017}

V549 Vel is also a nova detected in GeV gamma-rays \citep{li20hm}.
Figure \ref{optical_v5856_sgr_v906_car_v549_vel_v5855_sgr}c shows
the $V$ light curve of V549 Vel as well as a 0.98 $M_\sun$ WD 
(CO3) model light curve.
We assume that the $V$-band distance modulus is
$(m-M)_V= A_V + 5 \log (d / {\rm 10~pc})
= 3.1\times E(B-V) + 11.15 = 14.3$, where
we adopt $d=1.7$ kpc from Gaia EDR3 \citep{bai21rf}
and $E(B-V)=1.0$ mag from \citet{li20hm}. 
This nova experienced four jitters through maximum light
\citep[see Figure 2 of ][]{li20hm}.

We fit our photospheric $V$ (thick solid red line, labeled BB)
magnitudes with the first and second jitters and its decay.
For the last jitter (fourth), we apply the total $V$ 
(thick solid green, labeled TOTAL, i.e., BB+FF) light curve.
This is because free-free emission from plasma outside the photosphere
does not yet dominate the spectrum in the very early phase,
and we directly observed the photosphere \citep{li20hm}.
Free-free emission dominates the spectra only after 
a large amount of gas has been eventually ejected during the four jitters.

The first $V$ peak ($V=8.8$) is reached at MJD 58032.0 \citep[ANS
collaboration data denoted by the filled orange squares,][]{li20hm}
about 11 days before the second $V$ peak \citep[MJD 58043.4;][]{ayd20ci}.
We set $t=0$ day at this first $V$ peak (MJD 58032.0), although
\citet{ayd20ci} assumed that the second $V$ peak of $V=9.0$
was the optical maximum.

\citet{ayd20ci} obtained the velocity of the deepest narrow 
H$\alpha$ P Cygni absorption line,
from $-1000$ km s$^{-1}$ at day $-7.5$ (7.5 days before the optical maximum),
$-700$ km s$^{-1}$ at day $-2$, $-600$ km s$^{-1}$ at day $+3$, i.e.,
the velocity decreases toward the optical maximum,
and then increases to $-800$ km s$^{-1}$ at day $+53$. 
We adopt this velocity as the principal system.
Thus, we obtain the velocity of the principal system at post-maximum,
$v_{\rm p}\approx 800$ km s$^{-1}$.
On the other hand, we can see a shallow absorption trough between 
$-2700$ and $-2200$ km s$^{-1}$ in H$\alpha$ P Cygni profile on day
$+53$.  We attribute this absorption to the diffuse-enhanced system.
Then we have a velocity of $v_{\rm d}=(2700+2200)/2= 2450$ km s$^{-1}$
post-maximum.


The temperature just behind the shock is estimated to be 
k$T_{\rm sh}\approx ((2450 - 800)/1000)^2 = 2.7$ keV 
from equation (\ref{shock_kev_energy}). 
The thermal energy generation rate at the shock is given by 
equation (\ref{shocked_energy_generation}) and we obtain
$L_{\rm sh} \approx 1.8 \times 10^{37} {\rm erg~s}^{-1}
((2450 - 800)/1000)^3 (1000/2450) \times 0.5
= 1.7\times 10^{37}$ erg s$^{-1}$ for $\dot{M}_{\rm wind}=
0.5\times 10^{-4} ~M_\sun$ yr$^{-1}$ from fitting.

Hard X-rays were detected with Swift from 42 days after the 
maximum, and soft X-ray emission emerged from 222 days after the 
maximum \citep{li20hm, pag20bo}.
The plasma temperature of hard X-ray emission was estimated to be
1-2 keV, which is consistent with our analysis from the velocity
difference of $v_{\rm d} - v_{\rm p}$.

The emergence time of hard X-ray emission is estimated from
equation (\ref{hard_xray_emerging_time}).
\citet{li20hm} obtained a hydrogen column density of
$N_{\rm H}\sim 9.5^{+2.9}_{-2.7} \times 10^{21}$ cm$^{-2}$.
We obtain an ejecta mass of $M_{\rm ej}=1.1\times 10^{-5} ~M_\sun$
from our model light curve fitting in Figure
\ref{optical_v5856_sgr_v906_car_v549_vel_v5855_sgr}c.
The velocity of the shock is $v_{\rm shock}= v_{\rm p}= (600+800)/2= 700$
km s$^{-1}$ from the principal absorption system in V549 Vel
\citep{ayd20ci}.  Substituting these two values into equation
(\ref{hard_xray_emerging_time}), we have $t_{\rm H22} \sim 40$ day
for $N_{\rm H}\sim 10^{22}$ cm$^{-2}$.
This is broadly consistent with the hard X-ray emergence time
$t \sim 42$ days after the optical maximum \citep{li20hm}.

The photosphere shrinks, and the photospheric temperature increases
high enough to emit supersoft X-rays only after the optically thick winds stop
\citep{kat94h, hac06kb, hac10k}.  The wind duration is 
$t_{\rm wind}\sim 250$ days in our 0.98 $~M_\sun$ WD (CO3) model, being
broadly consistent with the emergence of soft X-ray emission
from 222 days after the maximum \citep{li20hm, pag20bo}.  The internal
absorption by the shocked ejecta is estimated from the hydrogen
column density $N_{\rm H}$ (equation (\ref{column_density_hydrogen_time}))
to be about $N_{\rm H}\sim 2\times 10^{20}$ cm$^{-2}$
at $t\sim 220$ day, low enough for soft X-rays to diffuse out.

GeV gamma-ray emissions were also detected with Fermi/LAT
from day $+5$  to day $+38$ \citep{li20hm}, that is, gamma-ray
emission appears several days after the optical maximum.
\citet{li20hm} pointed out that there is no clear correlation
between optical flux and gamma-ray flux.  Such uncorrelation
is a remarkable contrast to the excellent correlation in V5856 Sgr
and V906 Car \citep{li17mc, ayd20sc}. 
\citet{li20hm} also estimated the GeV gamma-ray flux to be
$L_\gamma \sim 4\times 10^{33}$ erg s$^{-1}$ for the distance
to the nova of $d=560$ pc.  If we adopt $d=1.7$ kpc from Gaia EDR3
\citep{bai21rf},
we have $L_\gamma \sim 4\times 10^{34}$ erg s$^{-1}$, about 10 times
larger, being consistent with those in the low $L_\gamma$ group
\citep[$\sim$ a few to several $10^{34}$ erg s$^{-1}$, e.g.,][]{li20hm}.

\citet{li20hm} obtained the ratio of $L_\gamma / L_{\rm opt} \approx 0.1$\%.
Then, we have $L_{\rm opt}\sim 4\times 10^{37}$ erg s$^{-1}$, which is
a half of the photospheric luminosity of our $0.98 ~M_\sun$ WD model,
i.e., $L_{\rm ph}\sim 8.5\times 10^{37}$ erg s$^{-1}$ at the optical maximum
\citep[$M_V\sim -5.5$, filled orange square in Figure
\ref{optical_v5856_sgr_v906_car_v549_vel_v5855_sgr}c, ][]{li20hm}.  
A majority of luminosity comes from the photosphere because
the shocked thermal energy generation rate (a few $10^{37}$ erg s$^{-1}$)
is smaller than the photospheric luminosity ($\sim 9 \times 10^{37}$ 
erg s$^{-1}$).  This is the reason why there is no clear correlation
between optical flux and GeV gamma-ray flux. 


Thus, our shock model based on a 0.98 $M_\sun$ WD (CO3) model
is consistent with the observation.  


\subsubsection{V5855 Sgr 2016}
\label{v5855_sgr_2016}

GeV gamma rays were detected in V5855 Sgr by Fermi/LAT \citep{nel19ml}.
Figure \ref{optical_v5856_sgr_v906_car_v549_vel_v5855_sgr}d
shows the $V$ light curve of V5855 Sgr and our 1.2 $M_\sun$ WD (CO3)
model light curve fitting.
Here, we assume the $V$-band distance modulus of
$(m-M)_V=A_V + 5 \log(d / {\rm 10~pc}) = 3.1\times E(B-V)
+ 14.23  = 15.5$.  Both the distance of $d=7.0$ kpc and reddening of
$E(B-V)=0.42$ mag are taken from \citet{mun17hf}.
Shifting horizontally our model $V$ light curve (green, labeled TOTAL),
we place the green line on the first $V$ peak ($V=7.6$).
The model $V$ light curve does not
follow the observation in the middle phase but approaches the 
observation in the last part.  This trend is very similar to those of
V5856 Sgr and V906 Car. 

\citet{ayd20ci} obtained the velocities of deepest narrow H$\alpha$
P Cygni absorption lines,
from $-250$ km s$^{-1}$ at day $-1.4$ (1.4 days before the optical maximum) to
$-600$ km s$^{-1}$ at day $+1.5$. 
They also showed a shallow broad absorption feature from $-3200$ 
to $-2500$ km s$^{-1}$ and a deepest narrow absorption at $-700$ km s$^{-1}$
on H$\alpha$ line at day $+7.2$.  
If we adopt $v_{\rm d}= (3200+2500)/2 = 2850$ km s$^{-1}$ and
$v_{\rm p}= 700$ km s$^{-1}$, the thermal temperature just behind
the shock is k$T_{\rm sh}= 2.1$ keV
from equation (\ref{shock_kev_energy}) and
the shocked thermal energy generation rate is 
$L_{\rm sh}= 1.8\times 10^{37} \times (2.15)^3/2.85 = 
6.3\times 10^{37}$ erg s$^{-1}$ for $\dot{M}_{\rm wind} =
1\times 10^{-4} ~M_\sun$ yr$^{-1}$, from equation
(\ref{shocked_energy_generation}).
The absolute $V$ magnitude at the peak is estimated to be
$M_V = m_V - (m-M)_V = 7.6 - 15.5 = -7.9$, which gives
the wind mass-loss rate of $\dot {M}_{\rm wind}= 
1.6\times 10^{-4} ~M_\sun$ yr$^{-1}$ 
and the total ejecta mass of $M_{\rm shell}= 0.35\times 10^{-5}~M_\sun$
on our 1.2 $M_\sun$ WD model \citep[see, e.g., ][]{hac20skhs}.      

\citet{nel19ml} obtained the GeV gamma-ray flux\footnote{In
their original text, $L_\gamma \sim 7\times 10^{35}
(d/4.5 {\rm ~kpc})$ erg s$^{-1}$. We add the power of 2
in the distance dependency.} 
of $L_\gamma \sim 7\times 10^{35} (d/4.5 {\rm ~kpc})^2$ erg s$^{-1}$,
7-15 days after discovery (discovery day $=$ JD 2457681.883).
If we adopt $d=7$ kpc from \citet{mun17hf}, this reads
$L_\gamma \sim 1.7\times 10^{36}$ erg s$^{-1}$, which is similar to
that of V5856 Sgr.  Then the conversion rate to gamma-ray
is about 2-3\%, i.e., $L_\gamma / L_{\rm sh}= 0.02$-0.03.


\citet{nel19ml} obtained hard X-ray emission from V5855 Sgr
with NuSTAR 15 days after discovery, but did not detect
four days later.  
Their analyzed results show that 
$N_{\rm H}= 2.2\times 10^{24}$ cm$^{-2}$,
k$T=$ 11 keV, $L_{\rm X}= 8\times 10^{35} 
(d/4.5 {\rm ~kpc})^2$ erg s$^{-1}$.  If we assume $d=7$ kpc,
this reads $L_{\rm X}= 2\times 10^{36}$ erg s$^{-1}$. 
We obtain the hydrogen column density of
$N_{\rm H}\approx 4\times 20^{22}$ cm$^{-2}$ 
from equation (\ref{column_density_hydrogen_time}), 
and the shock energy
generation rate of $L_{\rm sh}= 6\times 10^{37}$ erg s$^{-1}$
as calculated above.
Then the conversion rate to X-ray is about 3\%, i.e., 
$L_X / L_{\rm sh}= 0.03$.

The ratio of hard X-ray to gamma-ray is almost unity, i.e.,
$L_{\rm X} / L_\gamma \sim 1$, for this nova, which is very unusual
among various GeV gamma-ray detected novae because \citet{gor21ap}
showed a relation of $L_\gamma \sim (10^2$ - $10^3) L_{\rm X}$.   
We believe that this small ratio of X-rays to gamma rays,
$L_{\rm X}  / L_\gamma \sim 0.01$,
comes mainly from the quick decrease
in the wind mass-loss rate $\dot{M}_{\rm wind}$.
When hard X-ray emission emerges in a mid or later phase,
$\dot{M}_{\rm wind}$ had already decreased down to
$\lesssim 10^{-5} ~M_\sun$ yr$^{-1}$ or less. 
Thus, $L_{\rm sh}$ had decreased to much smaller than that of 
GeV gamma rays because $L_{\rm sh}$ is proportional to
$\dot{M}_{\rm wind}$.

Our 1.2 $M_\sun$ WD (CO3) model predicts the wind duration of
$t_{\rm wind}\sim 95$ days and hydrogen shell burning duration
of $t_{\rm Hburn}\sim 162$ days.  The so-called supersoft X-ray 
(k$T \lesssim 100$ eV) source phase starts after optically thick winds
stop \citep{hac06kb, hac10k} and ends after hydrogen shell burning stops.
Soft X-ray (0.3-1.0 keV) emission was once
observed with Swift 150 days after discovery \citep{gor21ap}. 
This epoch is in the period of supersoft X-ray source (SSS) phase
because optically thick winds had stopped in our model 
($t_{\rm wind}\sim 95$ days) but hydrogen is still burning 
($t_{\rm Hburn}\sim 162$ days).

\subsubsection{GeV Gamma-Ray Detectability in Novae}
\label{gev_gammaray_detectability}

We have analyzed several gamma-ray novae with rich optical information 
in previous sections. We may draw common properties for gamma-ray novae
having a red dwarf companion.
GeV gamma-ray emission could be detected with Fermi/LAT
in such a nova that satisfies
\begin{enumerate}
\item $v_{\rm d}-v_{\rm p}= v_{\rm wind}-v_{\rm shock} \gtrsim
1000$ km s$^{-1}$
\item $\dot {M}_{\rm wind} \gtrsim 1\times 10^{-4} ~M_\sun$ yr$^{-1}$
\item A relatively small distance to a nova, possibly
$d \lesssim 6-7$ kpc \citep[e.g.,][]{fra18jw, gor21ap}.
\end{enumerate}

\subsection{Emergence of Hard X-Rays in GeV Gamma-ray Novae}
\label{hard_x-ray_vs_gev_gamma-ray}

The above conditions for GeV gamma-ray emission are all necessary
conditions for hard X-ray detection.
We should further add a condition, that the hydrogen column density
decreases to $N_{\rm H} \sim 10^{21}$ cm$^{-1}$ before continuous wind stops.
If we define this time as $t_{\rm H21}$ from equation
(\ref{hard_xray_emerging_time}) and the wind stopping time
as $t_{\rm wind}$, this reads $ t_{\rm H21} \lesssim t_{\rm wind}$.
If we include the retarded (look back) time as discussed in section
\ref{section_hard_x-ray}, this correctly reads
$ t_{\rm H21} \lesssim t_{\rm wind} + t_{\rm ret}$,
where $t_{\rm ret} = (R_{\rm sh}(t) - R_{\rm ph}(t-t_{\rm ret}))
/ v_{\rm ph}(t-t_{\rm ret})$.  With another relation of
$(t_{\rm wind}+ t_{\rm ret})v_{\rm p} \sim v_{\rm d} t_{\rm ret}$,
the condition is rewritten as
\begin{equation}
t_{\rm H21} \lesssim t_{\rm wind}/(1 - {{v_{\rm p}}\over {v_{\rm d}}}).
\label{h21_wind_condition}
\end{equation}
We further simplify this condition to
\begin{equation}
t_{\rm H21} \lesssim 2 ~ t_{\rm wind}
\label{h21_wind_simplified_condition}
\end{equation}
with a broad relation of $v_{\rm d}\sim 2 v_{\rm p}$ \citep{mcl42}.
This also requires the observing epoch ($t_{\rm hX}$) of
\begin{equation}
t_{\rm H21} \lesssim t_{\rm hX} \lesssim  
t_{\rm wind}/(1 - {{v_{\rm p}}\over {v_{\rm d}}}),
\label{xobs_wind_condition}
\end{equation}
or simply,
\begin{equation}
t_{\rm H21} \lesssim t_{\rm hX} \lesssim 2 ~ t_{\rm wind},
\label{xobs_wind_simplified_condition}
\end{equation}
for hard X-ray detection.

\citet{gor21ap} discussed the relation between hard X-ray detected
novae and GeV gamma-ray detected novae.
Here, we simply assume that all GeV gamma-ray novae emit hard X-rays
and these hard X-rays could be detected 
if equation (\ref{xobs_wind_condition}) or equation 
(\ref{xobs_wind_simplified_condition}) is satisfied.
Actually, hard X-ray emission was observed in GeV gamma-ray novae,
mostly detected but two exceptions, V1324 Sco and V5856 Sgr.
No detection of hard X-rays are due either to the lack of
observation at proper $t_{\rm hX}$ in equation (\ref{xobs_wind_condition})
or to no proper $t_{\rm hX}$ times in equation (\ref{xobs_wind_condition}).
For example, if circumstellar or interstellar $N_{\rm H}$ is larger
than a few times $10^{22}$ cm$^{-2}$ or more, we obtain $t_{\rm H21}=\infty$, 
that is,
$t_{\rm H21} \gtrsim t_{\rm wind}/(1 - {{v_{\rm p}}\over {v_{\rm d}}})$
in equation (\ref{xobs_wind_condition}) and we have no detection of
hard X-rays.

\subsubsection{Observability of Hard X-Ray Emission}
\label{observability_hard_x-ray}

The condition (\ref{xobs_wind_condition})
means that hard X-rays are observable
only after the hydrogen column density decreases to
about $N_{\rm H}\sim 10^{21}$ cm$^{-1}$
but before the shock fades out, in other words, before the last wind 
collides with the shock.  In V1974 Cyg, we have $t_{\rm H21}= 61$ day
and $t_{\rm wind}= 250$ day from our 0.98 $M_\sun$ WD (CO3) model.
These values satisfy the condition, i.e., equations
(\ref{xobs_wind_condition}) or
(\ref{xobs_wind_simplified_condition}).

\subsubsection{Concurrent Detection of Hard and Supersoft X-Ray Emissions}
\label{concurrent_detection_hard_supersoft_x-ray}

The so-called supersoft X-ray (k$T \lesssim 100$ eV) source phase
of a nova starts after optically thick winds stop, that is,
$t \gtrsim t_{\rm wind}$ \citep[e.g.,][]{hac06kb, hac10k}.
It should be noted that hard X-ray emission is detected even
during the supersoft X-ray source phase because its duration extends
up to $\sim t_{\rm wind}/(1 - {{v_{\rm p}}\over {v_{\rm d}}})$ or
simply $\sim 2 ~t_{\rm wind}$.
In V959 Mon, hard X-ray emission was always observed
during the supersoft X-ray source phase \citep[e.g.,][]{pag13ow, hac18k}.

\subsubsection{V5856 Sgr 2016}
\label{no_hard_x-ray_detection_v5856_sgr}

We discuss the reason why no hard X-ray emission
was detected in V5856 Sgr.
The Swift satellite observed V5856 Sgr 
twice, $+1$ day and $+135$ day after the optical maximum. 
At the first observation the hydrogen column density was too large 
to detect hard X-rays. 
In our model the wind stopped at $t_{\rm wind}\sim 74$ day 
which gives $t_{\rm wind}/(1 - 1000/2200) = 136$ day in 
equation (\ref{xobs_wind_condition}). This is very close 
to the second observation epoch  $t_{\rm hX}\sim 135$ day.  
Thus, the wind could have stopped and shock has already disappeared, 
or even if the shock still survived, the shock luminosity 
has decreased as low as $L_{\rm sh}\lesssim 1\times 10^{34}$
erg s$^{-1}$, because the wind mass-loss rate decreased to 
$\dot {M}_{\rm wind}
\lesssim 1\times 10^{-7} ~M_\sun$ yr$^{-1}$. 

It has been argued that observed hard X-ray luminosity is orders of
magnitude smaller than that theory expects \citep{nel19ml, sok20mc}.
If we apply this argument to V5856 Sgr,
the upper limit of $L_X \lesssim 4\times 10^{32}$ erg s$^{-1}$
at +135 day obtained by \citet{li17mc} is consistent with
the upper limit on the shock model luminosity,
$L_{\rm sh} \lesssim 1\times 10^{34}$ erg s$^{-1}$ (+135 day) with
$L_{\rm X}/L_{\rm sh}\lesssim 0.03 L_{\rm X}/L_{\gamma} 
\lesssim 0.03\times 0.01 \sim 10^{-4}$.

Supersoft X-ray emission is observable only after the optically thick
winds stop.  This is because the photosphere shrinks, and the photospheric
temperature becomes high enough to emit supersoft ($< 100$ eV)
X-rays at a rate of $L_X\sim 10^{38}$ erg s$^{-1}$
\citep{oge93ok, kra96os, kah97h, kat97, hac06kb, hac10k, schw11no}.
In our 1.3 $M_\sun$ WD
model, hydrogen shell burning had ended at $+117$ day.
Thus, the second observation epoch ($+135$ day) with Swift
is possibly too late to detect both hard and soft X-rays.

\subsubsection{V1324 Sco 2012}
\label{no_hard_x-ray_detection_v1324_sco}

\citet{fin18cm} reported the P Cygni absorption profile on H$\alpha$ line
in V1324 Sco, 
the velocities of which are roughly measured to be 
 $-900$ km s$^{-1}$ (at $+3$ day, 2012 June 1st = JD 2456079.5 = $+0$ day),
 $-700$ km s$^{-1}$ ($+7$ day),
 $-600$ km s$^{-1}$ at the first optical maximum ($+13$ day),
 $-600$ km s$^{-1}$ ($+17$ day),
 $-800$ km s$^{-1}$ ($+19$ day).
Thus, we obtain the velocity of principal system, $v_{\rm p}\sim 800$
km s$^{-1}$, at post-maximum.
\citet{fin18cm} reported that the higher-velocity component appeared
a few weeks later and reached 2600 km s$^{-1}$.    They also reported
a slower component of 1000 km s$^{-1}$ at the same day.
GeV gamma rays were detected from $+14$ day ($+1$ day after the optical
maximum) to $+29$ day ($+16$ days after the maximum).
If we regard that $v_{\rm p}= 800$ km s$^{-1}$ and 
$v_{\rm d}= 2600$ km s$^{-1}$,
the shocked thermal temperature
is k$T_{\rm sh}= 3$ keV from equation (\ref{shock_kev_energy}),
which is high enough to emit hard (2-3 keV) X-rays.
The shocked thermal energy generation rate is estimated to be
$L_{\rm sh}= 1.8\times 10^{37} (1.8)^3/2.6 = 4\times 10^{37}$
erg s$^{-1}$ from equation (\ref{shocked_energy_generation})
for $\dot {M}_{\rm wind}= 1\times 10^{-4} ~M_\sun$ yr$^{-1}$.
    
 We obtain the ejecta mass of $M_{\rm ej}= 2.2\times 10^{-5} ~M_\sun$
from the light curve fitting with a $0.8 ~M_\sun$ WD
\citep[CO2;][]{hac19kb}.
The model predicts the wind duration of $t_{\rm wind}= 490$ day.
Equation (\ref{hard_xray_emerging_time}) gives $t_{\rm H21}\sim 140$ day.
The Swift satellite observed V1324 Sco two times, $+355$
and $+520$ day, so that these two observing epochs satisfy
the Equation (\ref{xobs_wind_condition}) or
(\ref{xobs_wind_simplified_condition}).
However, no hard or soft X-rays have been detected with Swift
\citep{fin18cm}.  

Soft X-rays are possibly not detected due to
a large interstellar extinction of $E(B-V)=1.32$ mag \citep[e.g.,][]{hac18kb,
hac19kb, hac21k}, corresponding to $N_{\rm H}=8.3\times 10^{21} \times
E(B-V)= 1.1 \times 10^{22}$ cm$^{-2}$ based on the relation of \citet{lis14}. 
In this sense, there may not be a proper time of $t_{\rm hX}$ satisfying
equation (\ref{xobs_wind_condition}) for hard X-rays because
always $N_{\rm H} \gtrsim$ a few times $10^{22}$ cm$^{-2}$.  

Moreover, this nova experienced a deep dust blackout,
as deep as 10 mag in the optical band, from $+45$ day to $+147$ day.
The last day of the dust blackout is close to our $t_{\rm H21}\sim 140$ day.
For a possible solution to this mystery, \citet{fin18cm} discussed that
dust is a signature of the thermal shock, which quickly cools down by emitting
high-energy (X-ray) photons that are absorbed by dense matter in the shell,
which are finally reemitted in optical or infrared \citep{der17ml}.
This may be supported by a relatively massive shell ($M_{\rm shell}=$
a few to several times
$10^{-5} ~M_\sun$) and slow expansion velocity of the shocked shell
($v_{\rm p} \lesssim 1000$ km s$^{-1}$).  For such a deep dust blackout
case, our simple $N_{\rm H}$ model of
equations (\ref{column_density_hydrogen_time}) and 
(\ref{hard_xray_emerging_time}) may not apply to V1324 Sco, at least,
after the dust blackout started.

On the other hand,
\citet{gor21ap} pointed out that no detection of hard X-rays in
V1324 Sco is due to its large distance ($d \gtrsim 6.5$ kpc).
The distance to V1324 Sco was estimated by \citet{hac18kb, hac19ka, hac21k}
to be $d=3.7\pm 0.4$ kpc together with the extinction of $E(B-V)=1.32$ mag
based on the color-magnitude diagram method and time-stretching method. 
The distance is also obtained by \citet{bai21rf} to be 
$d=4.9^{+1.9}_{-1.7}$ kpc based on the Gaia EDR3 parallax.
Thus, the distance to V1324 Sco is much smaller than 6.5 kpc, so we may 
exclude the reason of a large distance to V1324 Sco.

To summarize, we may conclude that no detection of hard X-rays is due
to a large absorption by dense matter of the shocked shell 
or high column density of the interstellar hydrogen, or both.
In other words, there is no proper $t_{\rm hX}$ time that satisfies
Equation (\ref{xobs_wind_condition}).

%

\subsection{Steady Versus Violent Mass Ejection}
\label{calm_violent_ejection}
A remarkable difference is found in the light curves of novae
between V382 Vel (Figure \ref{optical_mass_v382_vel_x45z02c15o20})
and V5856 Sgr
(Figure \ref{optical_v5856_sgr_v906_car_v549_vel_v5855_sgr}a).
The light curve of V382 Vel
well follows the 1.2 $M_\sun$ model light curve.
On the other hand, V5856 Sgr has a wavy plateau above the model
light curve, indicating violent vibration or multiple
ejections just after the optical maximum.
We dubbed the former case ``steady (mass) ejection''
and the latter case ``violent (mass) ejection.''
We do not know what makes the difference.   

\citet{kat09h, kat11h} proposed a hypothesis of transition between two types
of nova envelope solutions.  One is a hydrostatic solution without winds,
and the other is a steady-state wind solution.  There is a large
structure difference between them.  The static solution could be
perturbed by the companion star orbiting in the expanded WD envelope.
So the static envelope changes its configuration toward the steady-wind
solution \citep[][for more detail]{kat11h}.  Violent activities are
regarded as a relaxation process of transition from a static configuration
to a wind solution.  This can be applied to low-mass WDs
($M_{\rm WD} \lesssim 0.6$--0.7 $M_\sun$).

\citet{ayd20sc} proposed a complex density structured circumbinary torus
to explain multiple peaks for V906 Car.  Slow winds form circumstellar 
torus before the optical maximum, which has a complex density structure.
After that, fast nova winds slam into the multiple high-density regions
in the torus and makes multiple optical peaks.
These brightening processes occur far outside the photosphere and have no
effect on the outburst mechanism on the WD.  Therefore, the light curve 
eventually approaches the steady-state trend of the nova light curve,
as shown in Figures \ref{optical_v5856_sgr_v906_car_v549_vel_v5855_sgr}a,
\ref{optical_v5856_sgr_v906_car_v549_vel_v5855_sgr}b, and 
\ref{optical_v5856_sgr_v906_car_v549_vel_v5855_sgr}d, after the blast 
wave (forward shock) has broken out of the torus. 

On the other hand, V549 Vel could be a case of transition
between the two types of nova envelope solutions,
although the WD mass ($0.98 ~M_\sun$) is somewhat larger than
the mass range theoretically expected ($0.6-0.7 ~M_\sun$).
The first expansion seems to have no strong winds because its $V$ 
light curve almost follows the photospheric (labeled BB, solid red line)
luminosity in Figure \ref{optical_v5856_sgr_v906_car_v549_vel_v5855_sgr}c,
but the final outburst 31 days after the first outburst
seems to follow the total flux (of free-free (FF) emission
plus photospheric (BB) emission) light curve
(labeled TOTAL, solid green line = FF (blue) + BB (dashed red line)),
which shows strong winds.

We categorize a nova that shows a large excess in an early phase 
from our smooth decline model light curve as the group of violent ejection.
Many of the novae that are detected in GeV gamma rays are classified into
the group of violent ejection.  For example, V339 Del shows
a wavy structure similar to V5856 Sgr in the post-maximum phase
\citep[e.g.,][]{sko14dt}.  V1369 Cen, V5668 Sgr, V1324 Sco,
V5855 Sgr, V549 Vel, and V906 Car displayed multiple peaks 
\citep{che16js, fin18cm, ayd20ci}.

These hydrodynamical phenomena are difficult to be involved
in our 1D theoretical work because, in our time-dependent calculation,
the envelope approaches a steady state after the optical maximum,
and shows no signature of a wavy behavior.




\section{Conclusions}
\label{conclusions}

We propose a shock model in classical novae that can be responsible 
for absorption/emission line systems and hard X-ray emissions based on a
fully self-consistent nova outburst model calculated by \citet{kat22sha}.
Our main conclusions are as follows:\\

\noindent
{\bf 1.} 
We explained that a shock is inevitably generated outside the photosphere.
The wind velocity at the photosphere decreases with time before the optical
maximum but increases afterward. This is the result of the evolution
calculation by \citet{kat22sha}.
Simply assuming ballistic loci of winds outside the 
photosphere, we followed the motion of fluid (wind) and
obtained a strong shock formation soon after the optical maximum.
\\

\noindent
{\bf 2.} 
Increasing the wind velocity, from a few hundred to a thousand km s$^{-1}$,
results in strong compression of gas.  Thus, a strong shock (reverse shock)
arises soon after the optical maximum.
The velocity of the shocked matter ($v_{\rm shock}$)
is a few hundred km s$^{-1}$ smaller than that of the upstream wind
($v_{\rm wind}$), i.e., $v_{\rm wind} - v_{\rm shock}\sim (2-3)\times
100$ km s$^{-1}$. 
\\

\noindent
{\bf 3.}
A nova ejecta is divided by the shock into three parts;
the outermost expanding gas (earliest wind before maximum),
shocked shell, and inner fast wind.
These three parts are responsible for pre-maximum, principal,
and diffuse-enhanced absorption/emission line systems, respectively.
A large part of nova ejecta are eventually confined to the shocked shell.
The appearance of the principal system is consistent with the emergence
of the shock.  
\\

\noindent
{\bf 4.}
We interpret that the shock velocity $v_{\rm shock}$
is the velocity $v_{\rm p}$ of the principal system and
the inner wind velocity $v_{\rm wind}$ is the velocity $v_{\rm d}$
of the diffuse-enhanced system.
If we take the observed velocities of $v_{\rm d} - v_{\rm p}$ 
instead of our model values of $v_{\rm wind} - v_{\rm shock}$,
the shocked layer has a high temperature of
$k T_{\rm sh} \sim 1$ keV 
$((v_{\rm wind} - v_{\rm shock})/{\rm 1000 ~km~s}^{-1})^2
= 1$ keV $((v_{\rm d} - v_{\rm p})/{\rm 1000 ~km~s}^{-1})^2$,
where $v_{\rm d} - v_{\rm p}$ is the velocity difference
between the diffuse-enhanced ($v_{\rm d}$)
and principal ($v_{\rm p}$) systems.
This high temperature explains the hard X-ray emission from novae.
The shocked matter becomes as hot as a few times
$10^7$ K and emits thermal X-rays in the keV range, because
the velocity difference is usually $v_{\rm d} - v_{\rm p}
\gtrsim 1000$ km s$^{-1}$ in fast novae.
\\

\noindent
{\bf 5.} 
The shock arises after the optical maximum, so that hard (keV) X-ray emission
could be detected after the optical maximum if its optical depth becomes small.
\\

\noindent
{\bf 6.} 
Our shock model also explains GeV gamma-ray emissions from
classical novae based on Fermi acceleration of particles
(proton-proton collisions at the shock).
The shocked luminosity depends not only on the velocity difference,
$v_{\rm d}-v_{\rm p}$, but also on the wind mass-loss rate, 
$\dot{M}_{\rm wind}$. Because $\dot{M}_{\rm wind}$
is the largest at the optical maximum and quickly decreases afterward,
GeV gamma-ray emission is bright enough to be detected just
in the post-maximum phase. Then, it quickly decays with decreasing 
$\dot{M}_{\rm wind}$.
\\

\noindent
{\bf 7.} 
Thus, we strongly recommend high-resolution spectroscopic observations
of novae, in particular, before/after the optical maximum as well as
in the post-maximum phase.  These are essential to clarify
the intrinsic nature of shocks in nova ejecta.
\\

\begin{acknowledgments}
We thank 
the American Association of Variable Star Observers (AAVSO) and
the Variable Star Observers League of Japan (VSOLJ)
for the archival data of V1668 Cyg, V1974 Cyg, V382 Vel, V5855 Sgr,
V549 Vel, and V5856 Sgr.  We are also grateful to the anonymous referee
for his/her useful comments, which helped to improve the manuscript.
\end{acknowledgments}

\clearpage







\end{document}